\documentclass[a4paper,aps,preprintnumbers,showpacs,twocolumn,superscriptaddress,nofootinbib]{revtex4}

\pdfoutput=1
\usepackage{amsmath}
\usepackage{amssymb}
\usepackage{graphicx}
\usepackage{epsfig}
\usepackage{color}
\usepackage{url}
\usepackage{times}
\usepackage{bm}         
\usepackage{times}

\newcommand{\bea}{\begin{eqnarray}}
\newcommand{\eea}{\end{eqnarray}}
\newcommand{\beq}{\begin{equation}}
\newcommand{\eeq}{\end{equation}}

\newcommand{\cf}{\textit{cf.~}}
\newcommand{\ie}{\textit{i.e.,}~}
\newcommand{\eg}{\textit{e.g.,}~}
\newcommand{\ms}{{\rm ms}}
\newcommand{\km}{{\rm km}}
\newcommand{\hz}{{\rm Hz}}
\newcommand{\khz}{{\rm kHz}}
\newcommand{\mpc}{{\rm Mpc}}
\newcommand{\G}{\,{\rm G}}

\begin{document}


\title[Accurate evolutions of inspiralling and magnetized
  neutron-stars: equal-mass binaries]{Accurate evolutions of
  inspiralling and magnetized neutron-stars: equal-mass binaries}

\author{Bruno Giacomazzo}
\affiliation{Department of Astronomy, University of Maryland, College Park, Maryland, USA}
\affiliation{Gravitational Astrophysics Laboratory, NASA Goddard Space Flight Center, Greenbelt, Maryland, USA}
\affiliation{Max-Planck-Institut f\"ur Gravitationsphysik,
  Albert-Einstein-Institut, Potsdam Germany}

\author{Luciano Rezzolla}
\affiliation{Max-Planck-Institut f\"ur Gravitationsphysik,
  Albert-Einstein-Institut, Potsdam Germany}
\affiliation{Department of Physics and Astronomy, Louisiana State University, Baton Rouge, Louisiana, USA}

\author{Luca Baiotti}
\affiliation{Institute of Laser Engineering, Osaka University, Osaka, Japan}

\date{\today}

\begin{abstract}
  By performing new, long and numerically accurate
  general-relativistic simulations of magnetized, equal-mass
  neutron-star binaries, we investigate the role that realistic
  magnetic fields may have in the evolution of these systems. In
  particular, we study the evolution of the magnetic fields and show
  that they can influence the survival of the hypermassive neutron
  star produced at the merger by accelerating its collapse to a
  black hole. We also provide evidence that, even if purely
  poloidal initially, the magnetic fields produced in the tori
  surrounding the black hole have toroidal and poloidal components of
  equivalent strength. When estimating the possibility that magnetic
  fields could have an impact on the gravitational-wave signals
  emitted by these systems either during the inspiral or after the
  merger, we conclude that for realistic magnetic-field strengths $B
  \lesssim 10^{12}\G$ such effects could be detected, but only
  marginally, by detectors such as advanced LIGO or advanced
  Virgo. However, magnetically induced modifications could become
  detectable in the case of small-mass binaries and with the
  development of gravitational-wave detectors, such as the Einstein
  Telescope, with much higher sensitivities at frequencies larger than
  $\approx 2\,\khz$.
\end{abstract}

\maketitle

\section{Introduction} 
\label{sec:intro}

The use of improved and more accurate numerical techniques, together
with access to larger computational infrastructures, has brought
the simulation of binary neutron-star (BNS) systems to an
unprecedented level of maturity. A number of groups have reported on
calculations of BNSs with different levels of approximation, for equal-
and unequal-mass systems, with and without magnetic fields (see, \eg
\cite{Anderson2007:shortal, Baiotti08, Anderson2008, Baiotti:2009gk,
  Etienne08,Giacomazzo:2009mp, Kiuchi2009, Rezzolla:2010} for some of
the most recent works). Besides the obvious implications that these
systems have in our understanding of the origin of short $\gamma$-ray
bursts (GRBs), whose short rise times suggest that their central
sources have to be highly relativistic objects~\cite{Piran99},
BNS systems are expected to produce signals of amplitude
large enough to be relevant for Earth-based gravitational-wave (GW)
detectors and to be sufficiently frequent sources to be detectable
over the timescale in which the detectors are operative. The current
estimate for the detection rate relative to the first-generation
interferometric detectors is approximately 1 event per $40-300$ years,
increasing to an encouraging $10-100$ events per year for the
advanced detectors~\cite{Belczynski07}.

The detection of gravitational waves from neutron-star (NS) binaries
will also provide a wide variety of physical information on the
component stars~\cite{Andersson:2009yt}. This includes their mass,
spin, and radius, which would in turn provide vital clues on the
governing equation of state (EOS), and, possibly, their magnetic
field. However, for this information to be extracted it is essential
that accurate and long-term simulations are carried out, which span
the interval ranging from the early inspiral to the decaying tail of
the late ringing of the formed black hole (BH). This is indeed the
goal of this work, where we focus on whether or not present and future GW
detectors will be able to determine the level of magnetization
of NSs. This is not an academic question, as we know that NSs have very
large magnetic fields, and it is indeed via the magnetic-dipolar losses
that the vast majority of NSs are routinely detected as
pulsars~\cite{lorimer:lr}.  Yet, determining what the effects of
magnetic fields are on the inspiral and merger of BNSs is a remarkably
difficult task, requiring the solution of the Einstein equations
together with those of general-relativistic magnetohydrodynamics
(GRMHD). So far, only three GRMHD simulations of inspiralling BNSs
have been reported~\cite{Anderson2008,Etienne08,Giacomazzo:2009mp},
and while Refs.~\cite{Anderson2008,Etienne08} considered magnetic
fields that are astrophysically unrealistic\footnote{We note that
  although NSs with magnetic fields as large as $10^{16}$ are widely
  expected to be behind the phenomenology associated with magnetars,
  it is unrealistic to expect that the old NSs comprising the binary
  have magnetic fields that are so
  large.}~\cite{UGK98,Abdolrahimi2009}, only the work in
Ref.~\cite{Giacomazzo:2009mp} has studied magnetic fields of the
order of $\approx 10^{12}$ G, which are probably the strongest to be
expected for NSs near the merger. Ultralarge magnetic fields are,
however, not entirely uninteresting from a general-relativistic point
of view. Indeed, as discussed in~\cite{Giacomazzo:2009mp}, the
magnetic tension associated with these extremely large magnetic fields
can be so strong to reduce the stellar tidal deformations during the
inspiral and hence to lead to a slightly delayed time of merger.

Here we present a more extended analysis than the one given
in~\cite{Giacomazzo:2009mp} and report on a systematic investigation
of equal-mass BNSs systems through long-term simulations using the
highest resolutions to date. The calculations cover a range of magnetic fields from
$B\approx 10^{8}$ G up to $B\approx 10^{12}$ G, and two different
masses to distinguish the phenomenology of those binaries that lead to
a prompt collapse from those that lead instead to a delayed one (see
the discussion in~\cite{Baiotti08}). Overall, we find that magnetic
fields are amplified during the merger, when the turbulent motions,
triggered during the merger by the Kelvin-Helmholtz instability, curl
magnetic field lines producing a strong toroidal component that
reaches a strength comparable to the poloidal one. The toroidal field
maintains a value comparable or larger than the poloidal one during
the subsequent evolution of the hypermassive neutron star (HMNS)
formed after the merger. The stability of the latter, however, is
influenced by the strength of the poloidal field, which can transport
the angular momentum outwards and trigger the collapse of the HMNS to a
BH.  Furthermore, equipartition among the poloidal and toroidal
magnetic field components has been measured during the first $5$ ms
after the collapse of the HMNS, when the system consists of a rotating
BH surrounded by a massive, high-density torus.

We have also analyzed in detail the GW signal emitted by these systems
and found that for the timescales considered here, the overlaps in the
GWs between a nonmagnetized binary and a magnetized one are always
above what detectors such as Advanced LIGO (advLIGO) or Advanced Virgo
(advVirgo) can distinguish. Hence, it is very unlikely that present
detectors will be able to measure the presence of magnetic
fields. However, for sufficiently small-mass binaries, whose
corresponding HMNS could survive for up to a fraction of a second (see
the Appendix of~\cite{Rezzolla:2010}), the dephasing induced by the
presence of magnetic fields could be measurable, especially by those
detectors, such as the Einstein Telescope~\cite{Punturo:2010}, that
have higher sensitivities at frequencies larger than $\approx
2\,\khz$.

The paper is organized as follows. In
Sec.~\ref{sec:NumericalMethods} we first summarize the formalism we
adopt for the numerical solution of the Einstein and of the GRMHD
equations; we then describe briefly the numerical methods we
implemented in the {\tt Whisky} code~\cite{Baiotti03a, Baiotti04,
  Giacomazzo:2007ti}, we outline our mesh-refined grid setup, and we
finally describe the quasi-equilibrium initial data we use. In
Sec.~\ref{sec:bd} we describe the dynamics of the different models
by studying both the evolution of the matter and of the magnetic
field. In Sec.~\ref{sec:gw} we instead describe the GWs emitted by
these systems and we estimate the possibility to detect magnetic field
effects on those signals, while in Sec.~\ref{sec:conclusions} we
summarize our main results.

Here we use a spacelike signature $(-,+,+,+)$ and a system of units in
which $c=G=M_\odot=1$ (unless explicitly shown otherwise for
convenience).

\section{Mathematical and Numerical Setup}
\label{sec:NumericalMethods}

Most of the details on the mathematical and numerical setup used for
producing the results presented here are discussed in depth
in~\cite{Pollney:2007ss,Thornburg2003:AH-finding,Giacomazzo:2007ti,Giacomazzo:2009mp}. In
what follows, we limit ourselves to a brief overview and we describe
in more details only the main differences with respect to our previous
simulations.\\


\subsection{Einstein and Magnetohydrodynamics equations}
\label{sec:Einsten_MHD_eqs}

The evolution of the spacetime was obtained using the \texttt{Ccatie}
code, a three-dimensional finite-differencing code providing the
solution of a conformal traceless formulation of the Einstein
equations~\cite{Pollney:2007ss}. The GRMHD equations were instead
solved using the \texttt{Whisky}
code~\cite{Baiotti03a,Baiotti04,Giacomazzo:2007ti}, which adopts a
flux-conservative formulation of the equations as presented
in~\cite{Anton05} and high-resolution shock-capturing schemes
(HRSC). The {\tt Whisky} code implements several reconstruction
methods, such as Total-Variation-Diminishing (TVD) methods,
Essentially-Non-Oscillatory (ENO) methods~\cite{Harten87} and the
Piecewise Parabolic Method (PPM)~\cite{Colella84}. As already
discussed in~\cite{Giacomazzo:2009mp} the use of reconstruction
schemes of order high enough is fundamental for the accurate evolution
of these systems and in particular for assessing the impact of the
magnetic fields. Therefore all the results presented here have been
computed using the PPM reconstruction, while the Harten-Lax-van
Leer-Einfeldt (HLLE) approximate Riemann solver~\cite{Harten83} has
been used to compute the fluxes.

In order to guarantee the divergence-free character of the MHD
equations we have employed the flux-CD approach described
in~\cite{Toth2000}, but with one substantial difference, namely, that we use as an evolution
variable the vector potential instead of the magnetic field. In other
words, by using an expression similar to equation $(31)$
of~\cite{Toth2000}, we compute the electric field at the center of
each numerical cell by interpolating the fluxes computed at the
interfaces of the cell and then use it to evolve directly the vector
potential. We recall that in ideal MHD a relation exists between the
fluxes of the magnetic field $\vec{B}$ and the value of the electric
field $\vec{E}\equiv -\vec{\tilde{v}}\times\vec{\tilde{B}}$, where 
\begin{eqnarray}
&& \tilde{B}^i\equiv \sqrt{\gamma}B^i \,, \\
&& \tilde{v}^i\equiv \alpha v^i-\beta^i\,, 
\end{eqnarray}
and where $\gamma$ is the determinant of the $3$-metric, $v^i$ is the
$3$-velocity of the fluid as measured by an Eulerian observer,
$\alpha$ the lapse, and $\beta^i$ the shift vector. In particular, the
following relations hold in Cartesian coordinates
\begin{eqnarray}
&&  E_x = \tilde{F}^z(\tilde{B}^y)  = -\tilde{F}^y(\tilde{B}^z) \label{eq:Exflux} \; ,\\
&&  E_y = -\tilde{F}^z(\tilde{B}^x) = \tilde{F}^x(\tilde{B}^z) \label{eq:Eyflux} \; ,\\
&&  E_z = \tilde{F}^y(\tilde{B}^x)  = -\tilde{F}^x(\tilde{B}^y) \label{eq:Ezflux} \; ,
\end{eqnarray}
with
\begin{equation}
\tilde{F}^i(\tilde{B}^j)\equiv \tilde{v}^i\tilde{B}^j - \tilde{v}^j\tilde{B}^i\,.
\end{equation}
The evolution equations for the vector potential $\vec{A}$ and for the
magnetic field $\vec{B}$ can then be written as
\begin{eqnarray}
\partial_t \vec{A} &=& - \vec{E} \; , \label{eq:CTinduction} \\
\vec{\tilde{B}} &=& \vec{\nabla}\times \vec{A} \; .
\end{eqnarray}
Equation~\eqref{eq:CTinduction} is solved at the center of each cell
$(i,j,k)$, where the electric field is given by
\begin{widetext}
\begin{eqnarray}
E_x (x_i,y_j,z_k) &=& \frac{1}{4} \left( -\tilde{F}^y(\tilde{B}^z)_{(i,j+1/2,k)} - \tilde{F}^y(\tilde{B}^z)_{(i,j-1/2,k)} + \tilde{F}^z(\tilde{B}^y)_{(i,j,k+1/2)} + \tilde{F}^z(\tilde{B}^y)_{(i,j,k-1/2)} \right) \; ,\\
E_y (x_i,y_j,z_k) &=& \frac{1}{4} \left(\tilde{F}^x(\tilde{B}^z)_{(i+1/2,j,k)}+\tilde{F}^x(\tilde{B}^z)_{(i-1/2,j,k)} - \tilde{F}^z(\tilde{B}^x)_{(i,j,k+1/2)} - \tilde{F}^z(\tilde{B}^x)_{(i,j,k-1/2)} \right) \; ,\\
E_z (x_i,y_j,z_k) &=& \frac{1}{4} \left(-\tilde{F}^x(\tilde{B}^y)_{(i+1/2,j,k)}-\tilde{F}^x(\tilde{B}^y)_{(i-1/2,j,k)} + \tilde{F}^y(\tilde{B}^x)_{(i,j+1/2,k)} + \tilde{F}^y(\tilde{B}^x)_{(i,j-1/2,k)} \right) \; ,
\end{eqnarray}
\end{widetext}
$\tilde{F}^i(\tilde{B}^j)$ being the numerical flux computed at the
interface of the cell. 

Since the magnetic field is computed from the curl of the vector
potential using the same differential operator used to compute its
  divergence (\ie a central-difference scheme), its divergence free
character is guaranteed at essentially machine precision at all times,
also when using adaptive mesh-refinement (AMR). We note that a similar approach has been recently
implemented also in another code~\cite{Etienne:2010ui} and, in analogy
with~\cite{Etienne:2010ui}, we add a Kreiss--Oliger type of
dissipation~\cite{Kreiss73} to the evolution equation of the vector
potential in order to avoid the possible formation of spurious
post-shock oscillations in the magnetic-field evolution. It has indeed
been shown by~\cite{Rossmanith06} that applying TVD operators to the
vector potential does not guarantee automatically the TVD character of
the magnetic field, leading to possible post-shock oscillations in the
latter. The code has been validated against a series of tests in
special relativity~\cite{Giacomazzo:2005jy} and in full general
relativity (see~\cite{Giacomazzo:2007ti}).

The system of GRMHD equations is closed by an EOS and, as discussed in
detail in~\cite{Baiotti08}, the choice of the EOS plays a fundamental
role in the post-merger dynamics and significantly influences the
survival time against gravitational collapse of the HMNS produced by
the merger.

As already done in~\cite{Giacomazzo:2009mp}, also in this paper we have
employed the commonly used ``ideal-fluid'' EOS, in which the pressure
$p$ is expressed as $p = \rho\, \epsilon(\Gamma-1) $, where $\rho$ is
the rest-mass density, $\epsilon$ is the specific internal energy and
$\Gamma$ is the adiabatic exponent. Such an EOS, while simple,
provides a reasonable approximation and we expect that the use of
realistic EOSs would not change the main results of this work.


\begin{table*}[t]
  \caption{\label{table:ID}Properties of the eight equal-mass binaries
    considered: proper separation between the stellar centers
    $d/M_{_{\rm ADM}}$; baryon mass $M_{b}$ of each star; total ADM
    mass $M_{_{\rm ADM}}$; angular momentum $J$; initial orbital
    angular velocity $\Omega_0$; mean coordinate radius $r_e$ along
    the line connecting the two stars; ratio of the polar to the
    equatorial coordinate radii $r_p/r_e$; maximum rest-mass density
    $\rho_{\rm max}$; maximum initial magnetic field $B_{0}$, where $*$ 
    is $8, 10$ or $12$. Note that $M_{_{\rm ADM}}$ and $J$ are reported 
    as measured on the finite-difference grid.}
\begin{ruledtabular}
\begin{tabular}{lcccccccccc}
Binary &
\multicolumn{1}{c}{$d/M_{_{\rm ADM}}$} &
\multicolumn{1}{c}{$M_{b}~(M_{\odot})$} &
\multicolumn{1}{c}{$M_{_{\rm ADM}}~(M_{\odot})$} &
\multicolumn{1}{c}{$J~({\rm g\, cm^2/s})$} &
\multicolumn{1}{c}{$\Omega_0~({\rm rad/ms})$} &
\multicolumn{1}{c}{$r_e~({\rm km})$} &
\multicolumn{1}{c}{$r_p/r_e$}&
\multicolumn{1}{c}{$\rho_{\rm max}~({\rm gm/cm^3})$}& 
\multicolumn{1}{c}{$B_{0}~({\rm G})$} 
\\
\hline
\texttt{M1.45-B*}  & $14.4$ & $1.445$ & $2.680$ & $6.5084\times10^{49}$ & $1.78$ & $15.0 \pm 0.3$ & $0.899$ &
$4.58\times10^{14}$ & $0$ or $1.97 \times10^{*}$ \\
\texttt{M1.62-B*}  & $13.3$ & $1.625$ & $2.981$ & $7.7806\times10^{49}$ & $1.85$ & $13.6 \pm 0.3$ & $0.931$ & $5.91\times10^{14}$ & $0$ or $1.97 \times10^{*}$ \\
\end{tabular}
\end{ruledtabular}
\end{table*}

\subsection{Adaptive Mesh Refinements}
\label{sec:AMR}
Both the Einstein and the GRMHD equations are solved using the
vertex-centered AMR approach provided by
the \texttt{Carpet} driver~\cite{Schnetter-etal-03b}. Our rather basic
form of AMR consists in centering the highest-resolution level around
the peak in the rest-mass density of each star and in moving the
``boxes'' following the position of this maximum as the stars
orbit. The boxes are evolved as a single refinement level when they overlap.

The results presented below refer to simulations performed using $6$
levels of mesh refinement with the finest level having a resolution of
$h=0.1500\,M_{\odot}\simeq 221\,\mathrm{m}$. The grid structure is
such that the size of the finest grids is $24\,M_{\odot}\simeq
35.4\,\km$, while a single refinement level covers the region between
a distance $r=164\,M_{\odot}\simeq 242.2\,\km$ and
$r=254.4\,M_{\odot}\simeq 375.7\,\km$ from the center of the
domain. This region is the one in which our gravitational-wave
extraction is carried out, with a resolution of
$h=4.8\,M_{\odot}\simeq 7.1\,\km$ (as a comparison, the gravitational
wavelength is about $100\,\km$ and thus well-resolved on this
grid). In addition, a set of refined but fixed grids is set up at the
center of the computational domain so as to better capture the details
of the Kelvin-Helmholtz instability (\cf~\cite{Baiotti08}). Moreover,
after the merger, at about $8.5 \mathrm{ms}$, we enlarge the central
grid that is formed by the merging of the two initial boxes. We do
this in order to cover a cubical region with a side of about
$88.6\,\km$ and so better resolve not only the whole HMNS, but also
the BH-torus system which is produced by the collapse of the HMNS. For
all the simulations reported here we have used a reflection-symmetry
condition across the $z=0$ plane and a $\pi$-symmetry condition across
the $x=0$ plane\footnote{Stated differently, we evolve only the region
  $\{x\geq 0,\,z\geq 0\}$ applying a $180^{\circ}$-rotational-symmetry
  boundary condition across the plane at $x=0$.}. At the outer
boundary we instead used simple zeroth-order extrapolation on the MHD
variables (in practice, we just copy the value of the MHD quantities
from the outermost evolved point in each direction to the points of
the outer boundary in that direction). Also note that a very little
amount of matter and magnetic fields reaches the outer boundary, so
the effect of the outer-boundary conditions on the MHD and
hydrodynamical variables is negligible.

The timestep on each grid is set by the Courant condition (expressed
in terms of the speed of light) and so by the spatial grid resolution
for that level; the Courant coefficient is set to be $0.35$ on all
refinement levels.  The time evolution is carried out using
$4$th-order--accurate Runge-Kutta integration algorithm. Boundary data
for finer grids are calculated with spatial prolongation operators
employing $3$rd-order polynomials for the matter variables and
$5$th-order polynomials for the spacetime variables. The prolongation
in time employs $2$nd-order polynomials and this ensures a significant
memory saving, requiring only three timelevels to be stored, with
little loss of accuracy due to the long dynamical timescale relative
to the typical grid timestep.

The grid setup used here is therefore quite different from the one
adopted in our previous work on magnetized NS
binaries~\cite{Giacomazzo:2009mp}, where we used fixed
mesh-refinement in order to reduce the violation (generated by the
interpolation in the buffer zones) of the divergence-free constraint
of the magnetic field. Our current implementation, based of the
evolution of the vector potential, does not produce any violation of
the divergence-free condition of the magnetic field, since it
interpolates the vector potential instead of the magnetic field in the
buffer zones. Moreover, since the vector potential is stored at the
center of the cell, it is possible to use without modification the
prolongation and restriction operators currently available in the
\texttt{Carpet} driver. This makes it possible to use the moving-grid
setup that has been utilized with success in our previous
general-relativistic hydrodynamics simulations.

\subsection{Initial data}
\label{sec:initial_data}

The initial data are the same as those used
in~\cite{Baiotti08,Giacomazzo:2009mp}.  They were produced by
Taniguchi and Gourgoulhon~\cite{Taniguchi02b} with the multi-domain
spectral-method code {\tt LORENE}~\cite{lorene}. The initial solutions
for the binaries are obtained assuming a quasi-circular orbit, an
irrotational fluid-velocity field, and a conformally-flat spatial
metric. The matter is modeled using a polytropic EOS $p = K
\rho^{\Gamma}$ with $K=123.6$ and $\Gamma=2$, in which case the
maximum gravitational mass is $M_{_{\rm ADM}}\simeq 1.82\,M_{\odot}$
for a nonrotating star and $M_{_{\rm ADM}}\simeq 2.09\,M_{\odot}$ for
a uniformly rotating one. Since no self-consistent solution is
available for magnetized binaries yet, a poloidal magnetic field is
added a-posteriori using the vector potential
\begin{equation}
\label{eq:Avec}
A_{\phi} \equiv \varpi^2 A_b\, {\rm max}\,(p-p_{\rm cut},0)^{n_{\rm s}} \,,
\end{equation}
where $\varpi \equiv \sqrt{x^2+y^2}$, $A_b>0$ parameterizes the
strength of the magnetic field, $p_{\rm cut}$ defines where in the NS
the magnetic field goes to zero, and $n_{\rm s}$ determines the degree
of differentiability of the potential. The components of the magnetic
field are then computed by taking the curl of the Cartesian components
of Eq.~(\ref{eq:Avec}) to enforce that the divergence of the magnetic
field is zero at machine precision. Here we have set $p_{\rm
  cut}=0.04\, {\rm max}(P)$, and $n_{\rm s}=2$ to enforce that both
the magnetic field and its first derivative are zero at $p=p_{\rm
  cut}$. In Ref.~\cite{Anderson2008} the magnetic field was built with
an expression equivalent to~\eqref{eq:Avec}, but with $p_{\rm cut}$
set to the pressure in the atmosphere, and in Ref.~\cite{Etienne08}
the expression used is slightly different and $P_{\rm cut}$ is set to
be $4\%-0.1\%$ of ${\rm max}(p)$; in both Refs.~\cite{Anderson2008}
and~\cite{Etienne08} $n_{\rm s}=1$.

Table~\ref{table:ID} lists some of the properties of the eight
equal-mass binaries considered here. More specifically, we have
considered two classes of binaries differing in the initial masses,
\ie binaries \texttt{M1.45-B*}, and binaries \texttt{M1.62-B*}. For
each of these classes we have considered four different magnetizations
(indicated by the asterisk) so that, for instance, \texttt{M1.45-B12}
is a low-mass binary with a maximum initial magnetic field $B_{\rm
  0}=1.97 \times 10^{12}\,\G$. Note that the binaries with zero
magnetic fields are the same as those evolved in
Ref.~\cite{Baiotti08}.

\subsection{Gravitational-Wave Extraction}
\label{sec:GWs_setup}

Details about the algorithms implemented in the code to extract the GW
signal can be found in~\cite{Baiotti08}. Here we just remind the
reader that we compute the waveforms using two different methods. The
first one is based on the Newman-Penrose formalism and computes the
Weyl scalar $\Psi_4$. The gravitational-wave polarization amplitudes
$h_+$ and $h_\times$ are then related to $\Psi_4$ by simple time
integrals~\cite{Teukolsky73}
\begin{equation}
\ddot{h}_+ - {\rm i}\ddot{h}_{\times}=\Psi_4 \ ,
\label{eq:psi4_h}
\end{equation}
where the double overdot stands for the second-order time derivative.

The second method is instead based on the measurements of the
nonspherical gauge-invariant perturbations of a Schwarzschild BH (see
refs.~\cite{Abrahams97a,Rupright98,Rezzolla99a} for some applications
of this method to Cartesian-coordinate grids). In practice, a set of
``observers'' is placed on $2$-spheres of fixed radius where we
extract the gauge-invariant, odd-parity (or {\it axial}) current
multipoles $Q_{\ell m}^\times$ and even-parity (or {\it polar}) mass
multipoles $Q_{\ell m}^+$ of the metric
perturbation~\cite{Moncrief74,Abrahams95b}. The $Q^+_{\ell m}$ and
$Q^\times_{\ell m}$ variables are related to $h_+$ and $h_\times$
as~\cite{Nagar05}
\begin{equation}
\label{eq:wave_gi}
h_+-{\rm i}h_{\times} =
  \dfrac{1}{\sqrt{2}r}\sum_{\ell,\,m}
  \Biggl( Q_{\ell m}^+ -{\rm i}\int_{-\infty}^t Q^\times_{\ell
          m}(t')dt' \Biggr)\,_{-2}Y^{\ell m}\ .
\end{equation}
Here $_{-2}Y^{\ell m}$ are the $s=-2$ spin-weighted spherical
harmonics and $(\ell, m)$ are the indices of the angular
decomposition.

Since the two methods have been shown to give waveforms that are
identical up to the truncation error, we will here use $h_+$ computed
only with the gauge-invariant quantities and we will focus only on the
$\ell=2, m=2$ mode since the others have amplitudes which are
negligible compared to this. All the waveforms have been extracted at a
radius $r_{\rm iso}=200M_{\odot}\approx 300\,\km$. We also ignored
the contribution from the spherical harmonics since they depend on the
direction of the source with respect to the detector and contribute as
a multiplication factor of order $1$; thus they do not modify the results
presented here.

\begin{figure*}
  \begin{center}
    \includegraphics[angle=0,width=8.0cm]{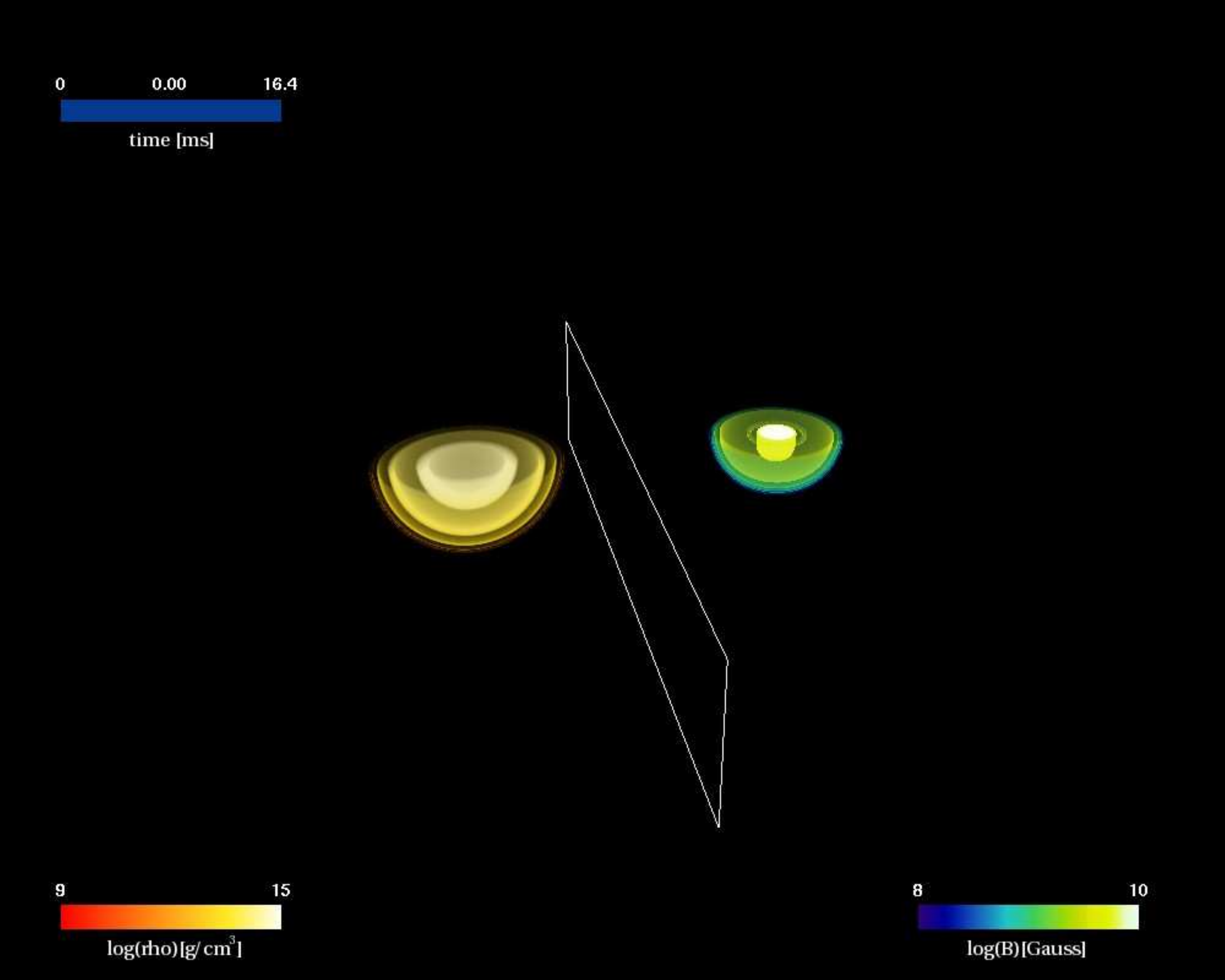}
    \hskip 0.5cm
    \includegraphics[angle=0,width=8.0cm]{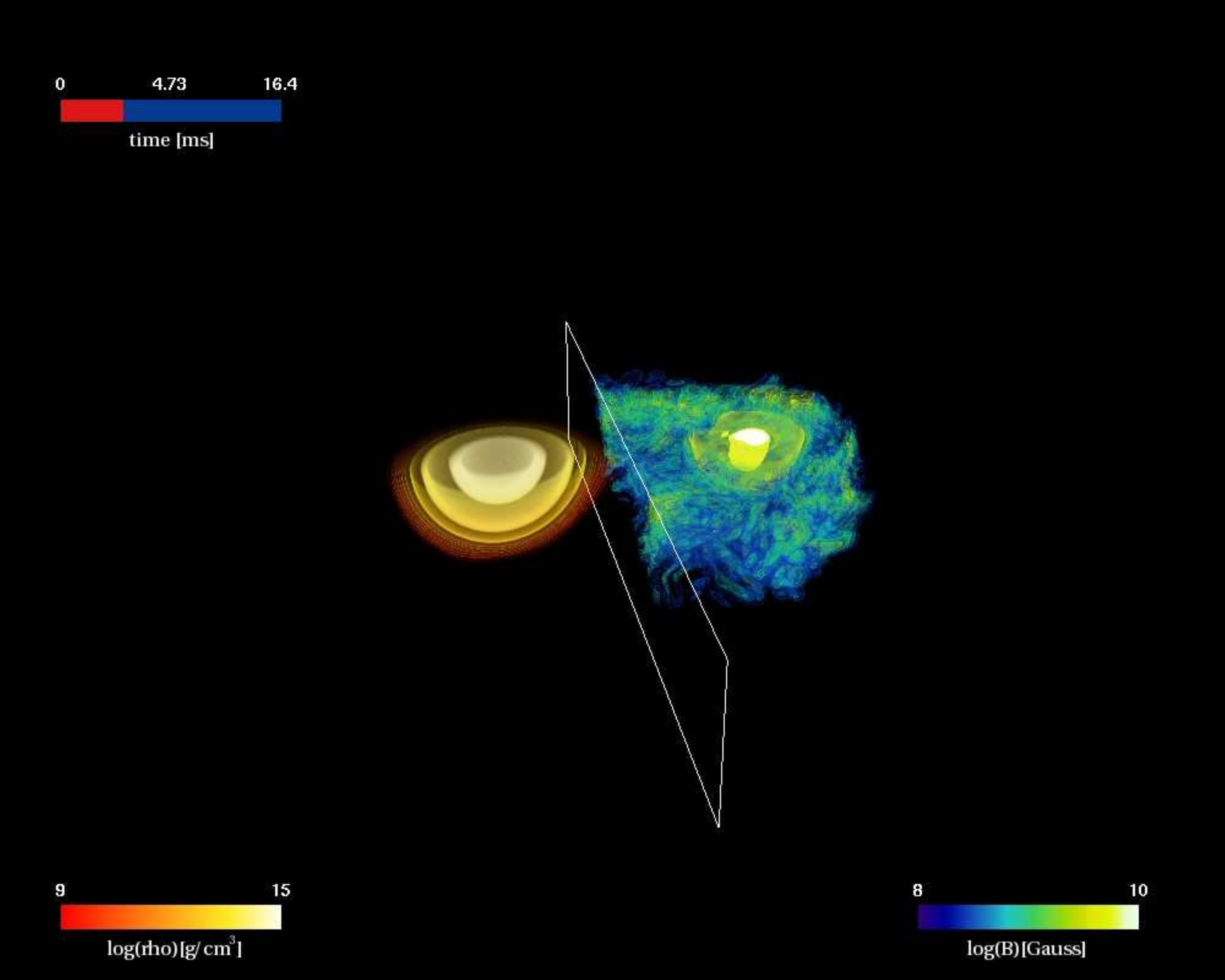}
    \vskip 0.5cm
    \includegraphics[angle=0,width=8.0cm]{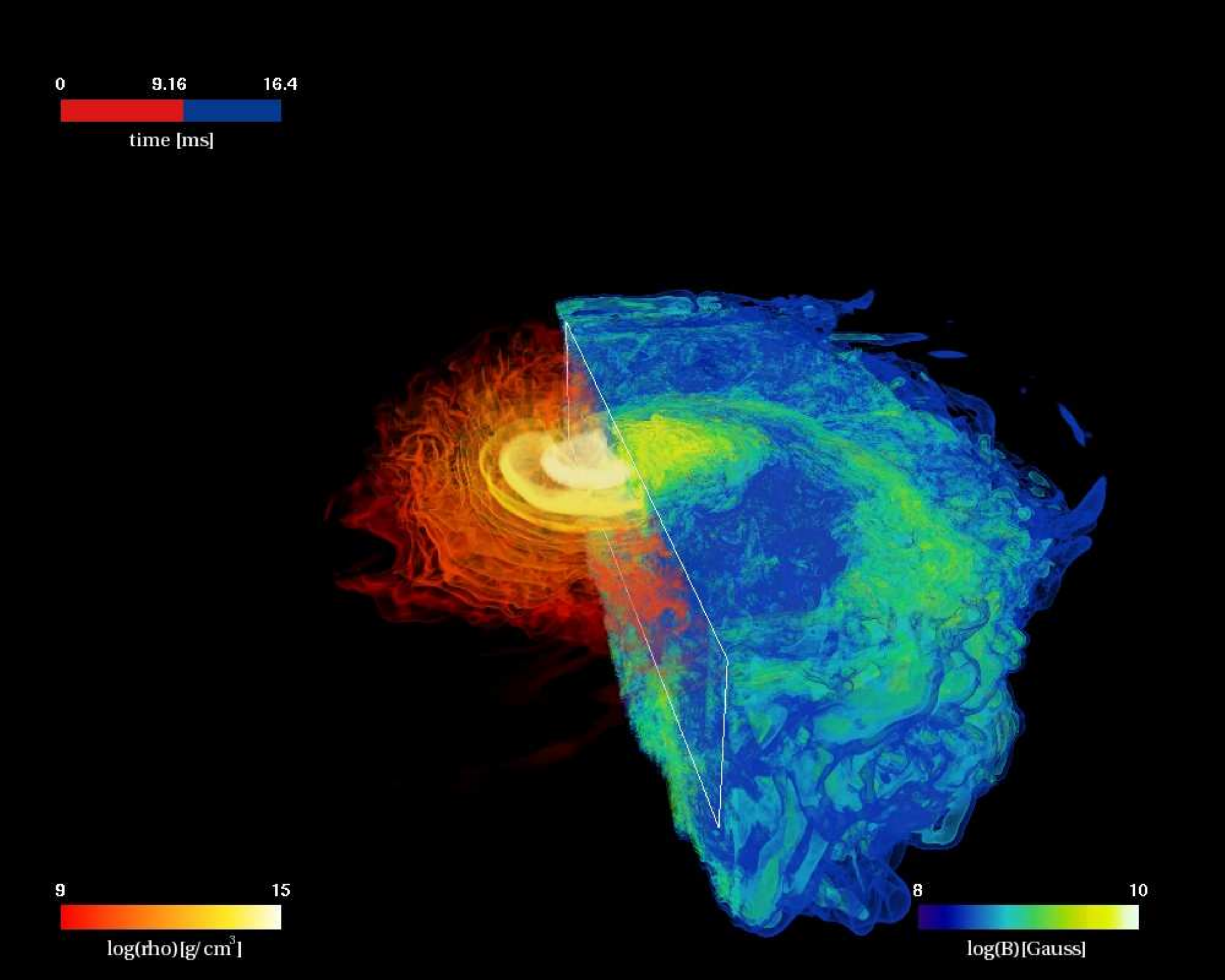}
    \hskip 0.5cm
    \includegraphics[angle=0,width=8.0cm]{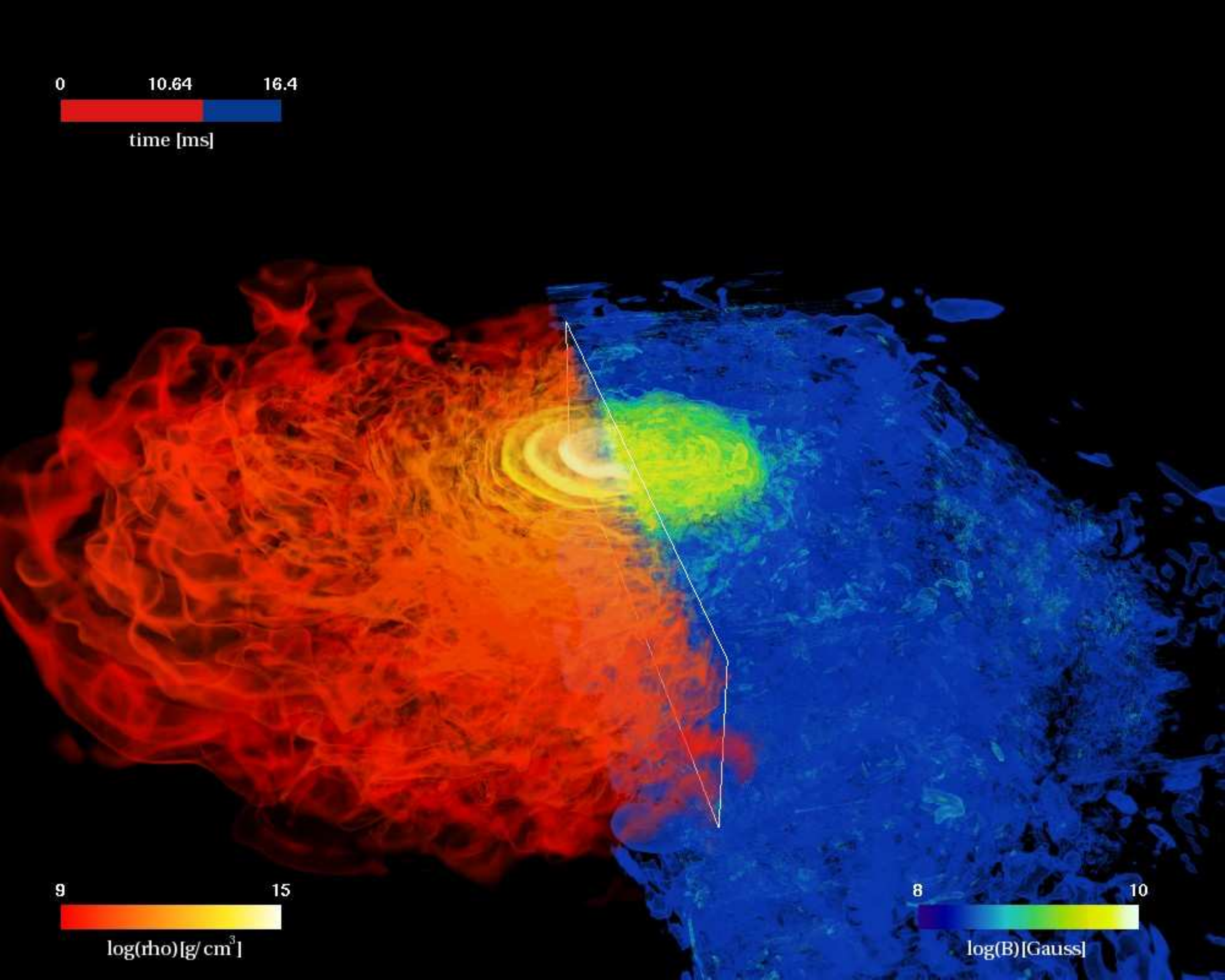}
    \vskip 0.5cm
    \includegraphics[angle=0,width=8.0cm]{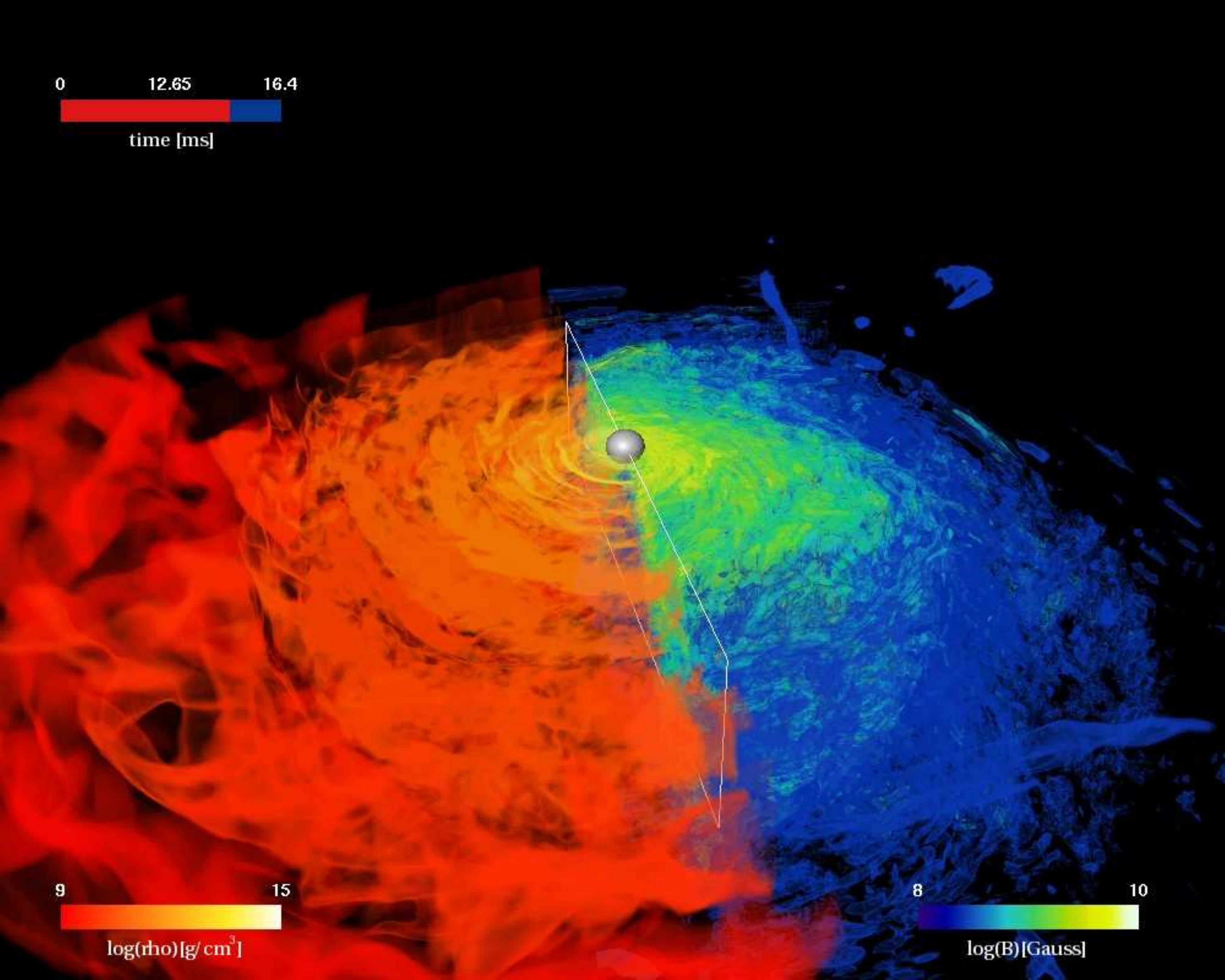}
    \hskip 0.5cm
    \includegraphics[angle=0,width=8.0cm]{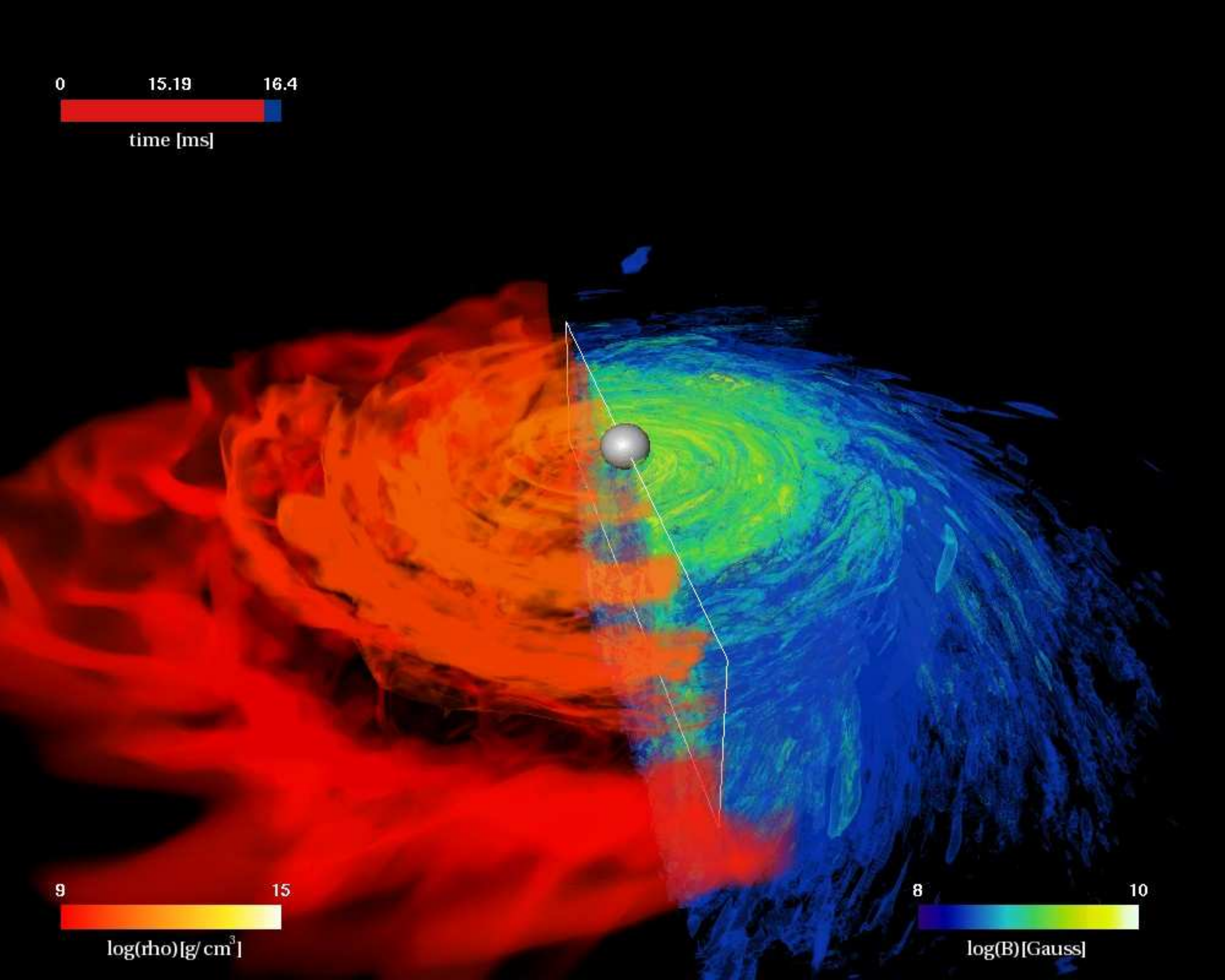}
  \end{center}
    \vskip 0.5 cm
  \caption{\label{fig0.5}{Snapshots at representative times of the
      evolution of the high-mass binary with initial maximum magnetic
      field of $10^{10}\,{\rm G}$, \ie \texttt{M1.62-B10}. Shown with
      two different color-code maps are the rest-mass density $\rho$
      (red-yellow) and the magnetic field $|B|$
      (blue-green-yellow-white). To better visualize the inner
      structure we plot only the values on $z<0$. In order to show the
      two scalar quantities at the same time, they are shown on either
      side of a fictitious screen ($\rho$ on the left and $|B|$ on the
      right). The first four panels refer respectively to the binary
      at the initial separation of $45$ km, to the binary after two
      orbits, to the merger and to the bar-deformed HMNS. The last two
      panels, instead, refer respectively to when the BH has just been
      formed and to a subsequent stage of the quasi-stationary
      evolution of the BH-torus system. The grey spheroidal surface in
      the center represents the location of the apparent horizon.}}
\end{figure*}

\subsection{Accuracy of the Results}
\label{sec:accuracy}

A reliable assessment of the truncation error is essential to draw
robust conclusions on the results of numerical simulations. Following
a procedure discussed in detail in Ref.~\cite{Baiotti:2009gk}, also
here we have carried out a systematic measurement of the accuracy and
convergence properties of our simulations, and deduced a corresponding
\textit{``error-budget''}. The main conclusions are very similar to
those drawn in Ref.~\cite{Baiotti:2009gk}, which for compactness only
we briefly recall here. More specifically, we showed that with typical
(finest) resolutions of $h\simeq 0.12\,M_{\odot}- 0.19\,M_{\odot}$, the
results show the expected convergence rate of $1.8$ during the
inspiral phase, which however drops to $1.2$ at the merger and during
the evolution of the HMNS. This deterioration of the convergence rate
is due mostly to the strong shocks which form during the merger and
which HRSC schemes can reproduce at $1$st-order only. Furthermore,
physical quantities, such as the rest-mass, are conserved with a
relative error of $\lesssim 10^{-6}$, while the energy and the angular
momentum are conserved to $\lesssim 1\%$ after taking into account the
parts lost to radiation. Finally, the expected agreement in both phase
and amplitude is found in the waveforms extracted from different
detectors within the same simulation or from the same detector but at
different resolutions. Such waveforms have been found to be also
convergent at a rate of $1.8$ (see~\cite{Baiotti:2009gk} for
details). Finally, for all the simulations reported here the violation
of the Hamiltonian constraint has an L2-norm which is $\lesssim
10^{-4}/M_{\text{ADM}}^2$ for the high-mass binaries and $\lesssim
10^{-5}/M_{\text{ADM}}^2$ for the low-mass ones, for which no BH is
  formed.

\section{Binary dynamics}
\label{sec:bd}

As mentioned above, in order to highlight some of the most salient
aspects of the binary dynamics it will be sufficient to consider two
main classes of initial configurations: \texttt{M1.62-B*} and \texttt{1.45-B*}. These models
differ only in the mass, the first being composed of stars each having
a rest mass of $1.625\,M_{\odot}$ (which we refer to as the
\textit{``high-mass binaries''}), the second of stars of rest mass
$1.445\,M_{\odot}$ (which we refer to as the \textit{``low-mass
  binaries''}). The use of these two classes is useful to distinguish
the phenomenology of binaries whose merger leads to a prompt collapse
of the HMNS from those where the HMNS can instead survive for
several tens of milliseconds and up to a fraction of a second (see the
discussion in~\cite{Baiotti08}). We also note that in the case of the
unmagnetized models, the dynamics is the same as the ones described
in~\cite{Baiotti08}, to which we refer the interested reader for a
more detailed description of the evolution of the matter and of the
hydrodynamical instabilities such as the Kelvin-Helmholtz instability.

A synthetic overview of the dynamics is summarized in
Fig.~\ref{fig0.5}, which shows snapshots at representative times of
the evolution of the high-mass binary with an initial maximum magnetic
field of $10^{10}\,{\rm G}$, \ie \texttt{M1.62-B10}. Shown with two
different color-code maps are the rest-mass density $\rho$
(red-yellow) and the magnetic field $|B|$ (blue-green-white). To
better visualize the inner structure we plot only the values on $z<0$. In
order to show the two scalar quantities at the same time, they are
shown on either side of a fictitious screen ($\rho$ on the left and
$|B|$ on the right). The first four panels refer respectively to the
binary at the initial separation of $45$ km ($t=0\,\ms$), to the
binary after two orbits ($t=4.7\,\ms$), to the merger ($t=9.2\,\ms$)
and to the bar-deformed HMNS ($t=10.6\,\ms$). The last two panels,
instead, refer respectively to when the BH has just been formed
($t=12.6\,\ms$) and to a subsequent stage of the quasi-stationary
evolution of the BH-torus system ($t=15.2\,\ms$).

\begin{figure*}
  \begin{center}
    \includegraphics[angle=0,width=7.0cm]{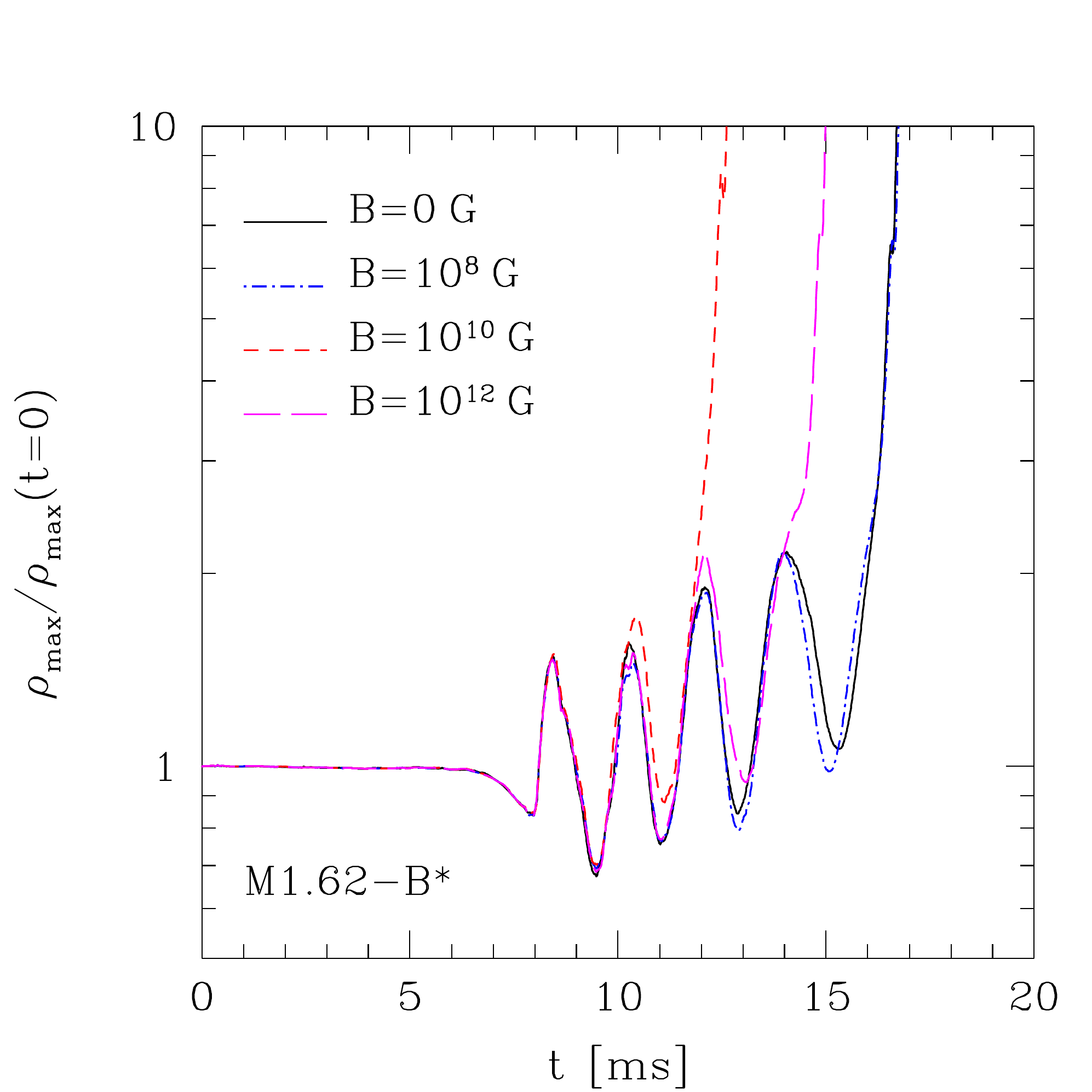}
    \includegraphics[angle=0,width=7.0cm]{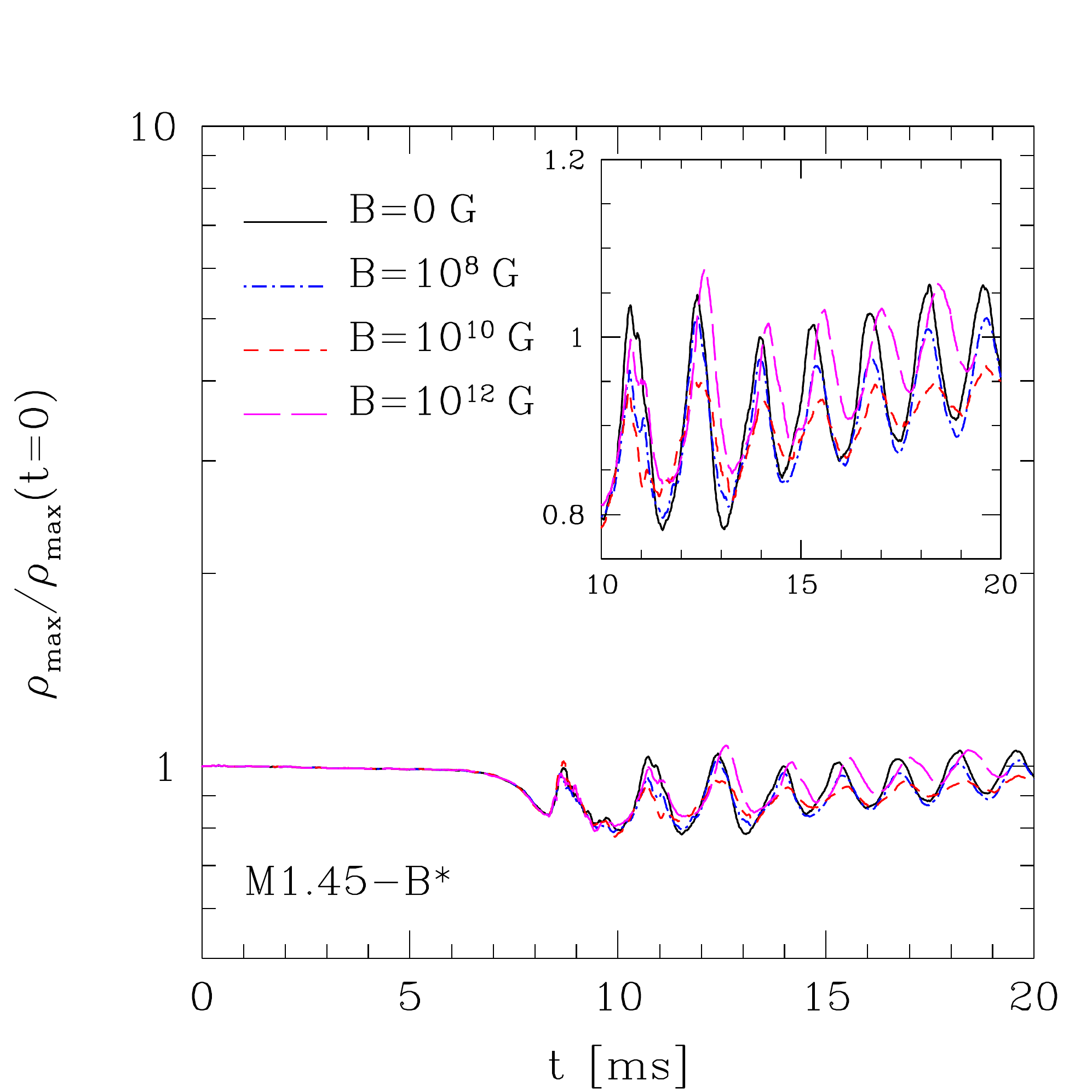}
    \includegraphics[angle=0,width=7.0cm]{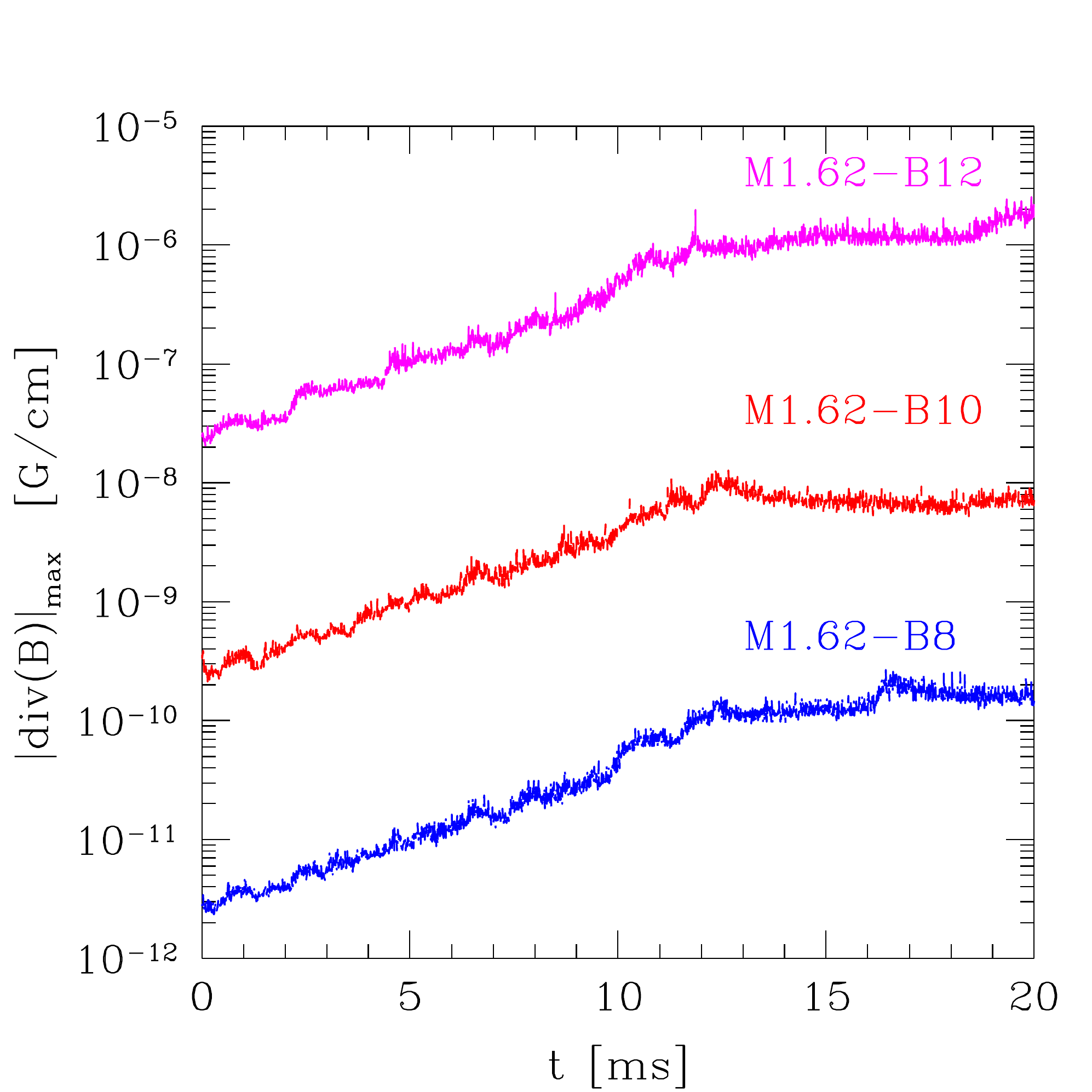}
    \includegraphics[angle=0,width=7.0cm]{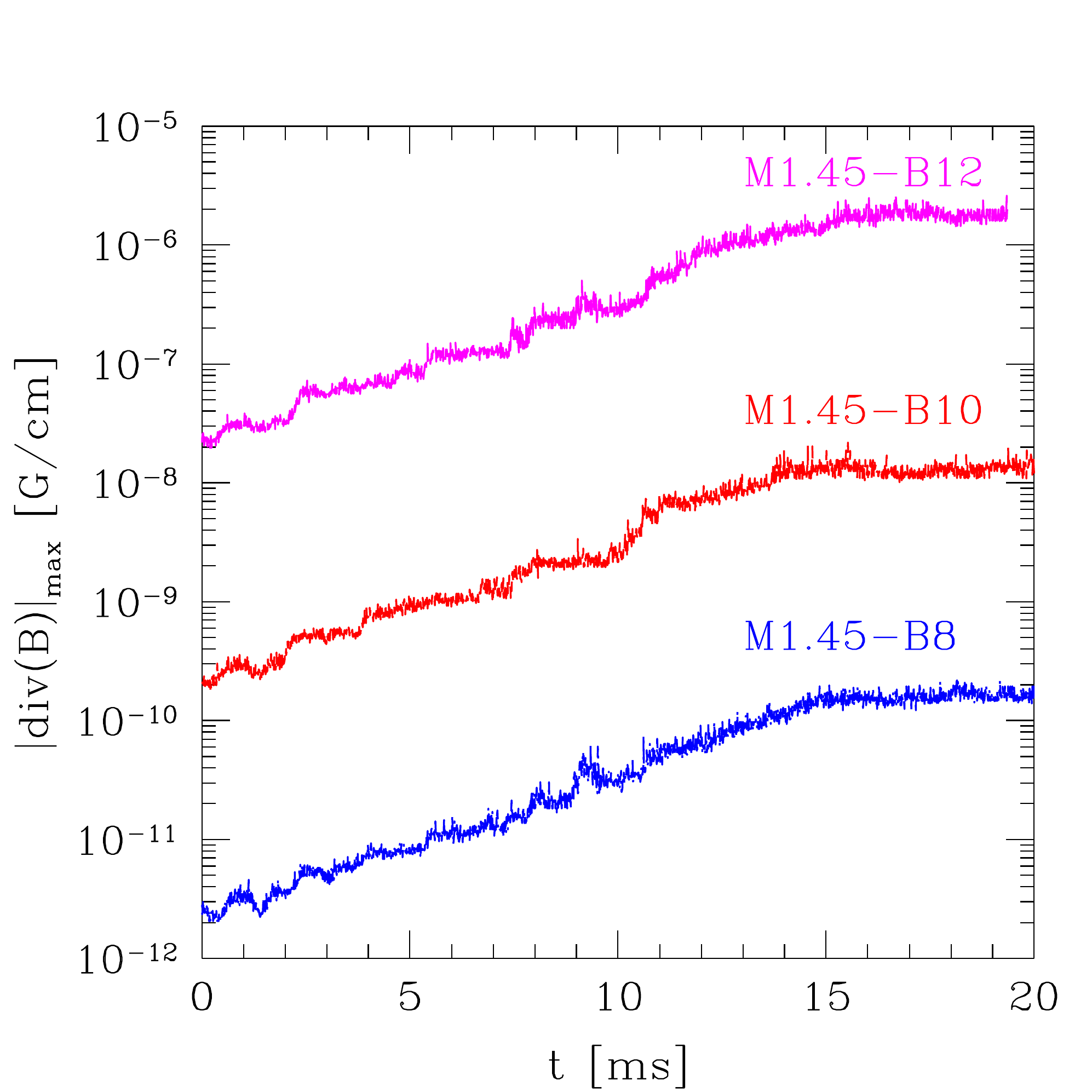}
    \includegraphics[angle=0,width=7.0cm]{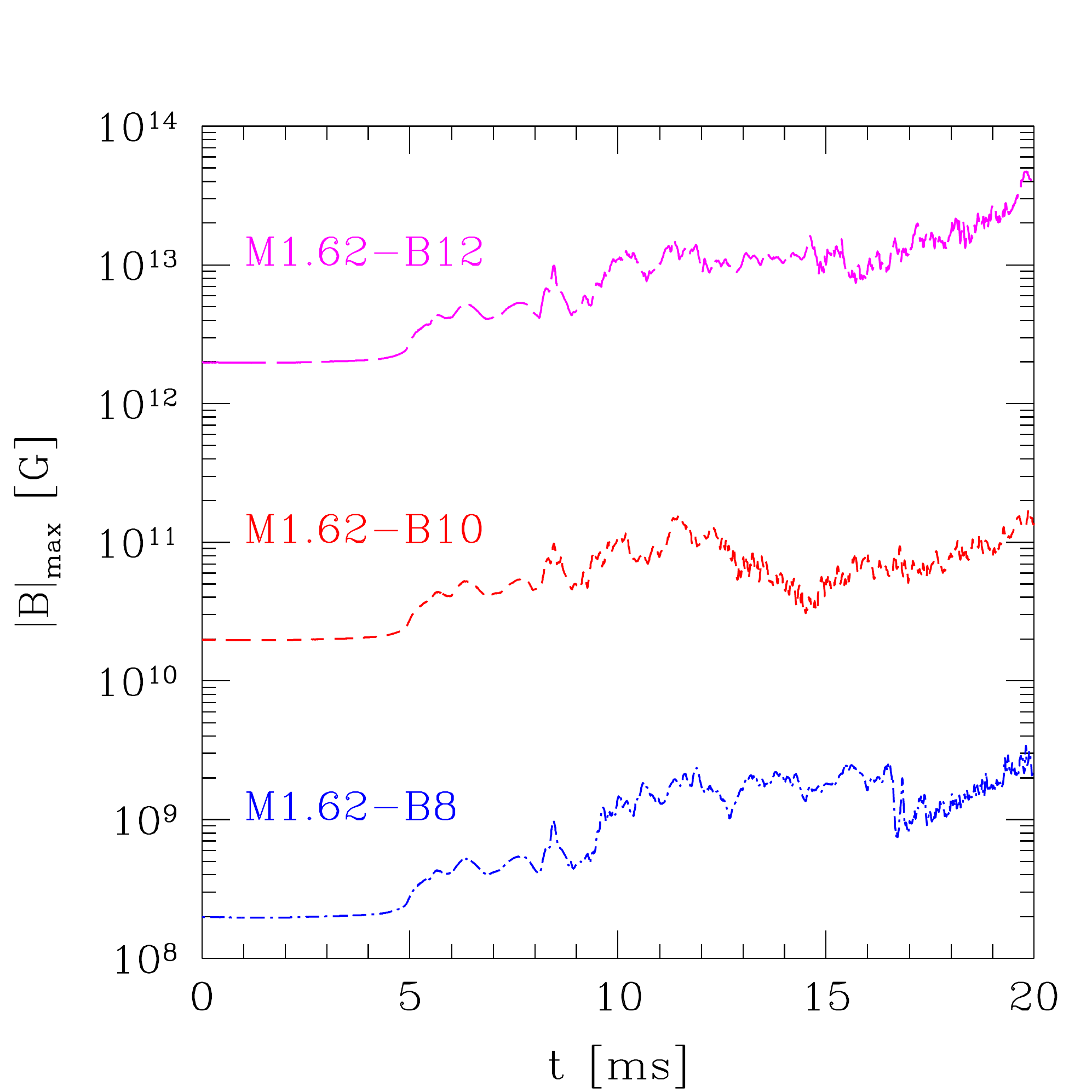}
    \includegraphics[angle=0,width=7.0cm]{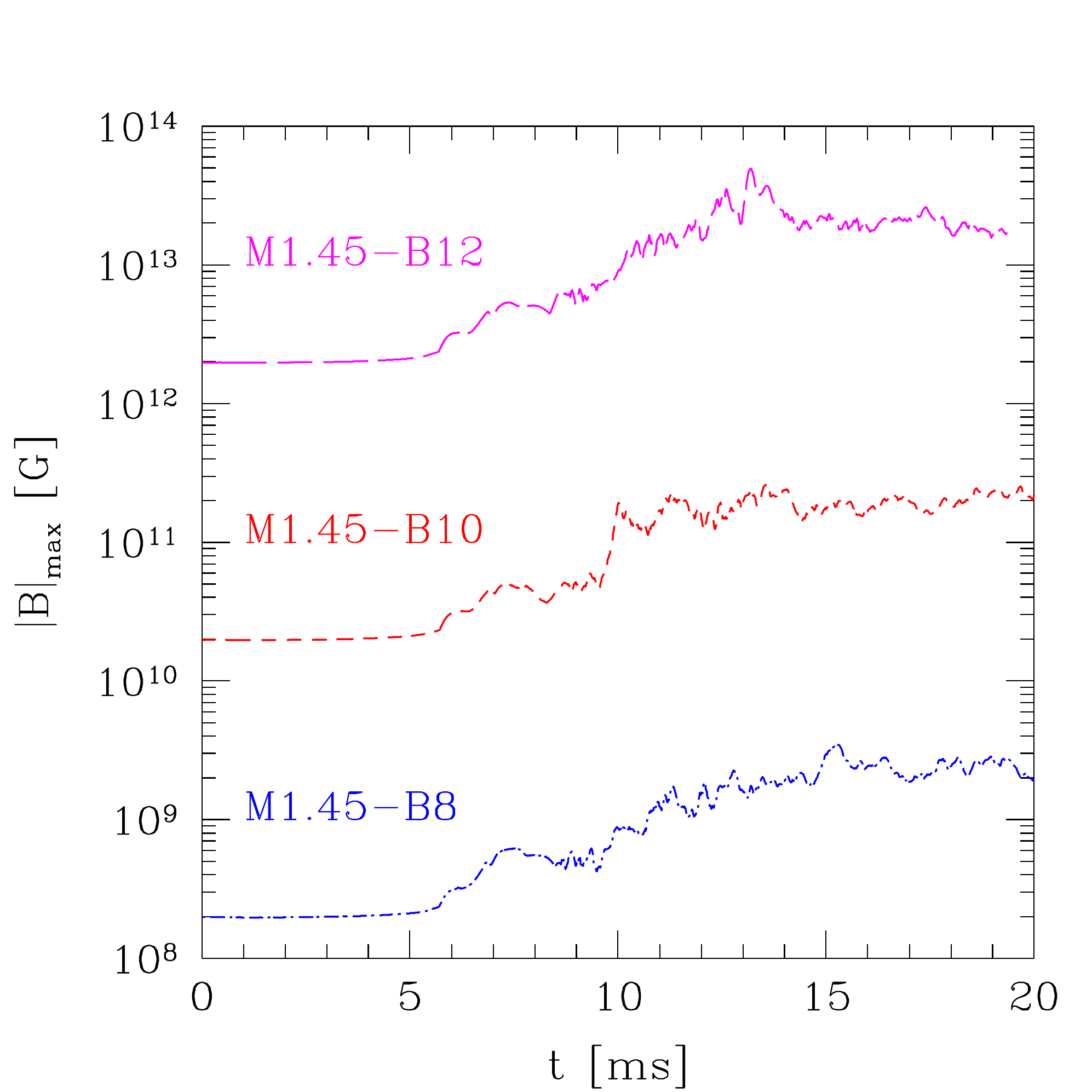}
  \end{center}
    \vskip -0.5 cm
  \caption{\label{fig1}Evolution of the maximum of the rest-mass
    density $\rho$ normalized to its initial value (top row), of the
    maximum of the absolute value of the divergence of $B$ (middle
    row), and of the maximum of the magnetic field strength $|B|$
    (bottom row). The left and right columns refer to the high-mass
    and low-mass binaries, respectively. Note that in the case of the
    high-mass models (left column), the values of $|B_{\rm max}|$ after
    BH formation refer, for the large majority of the time, to matter
    outside the apparent horizon and in the torus.}
\end{figure*}

With this overall qualitative behavior of the binary in mind, we will
next consider a more quantitative discussion of the evolution of the
magnetic fields and we will only briefly summarize the dynamics of the
matter. In doing this we will present in Figs.~\ref{fig1}
and~\ref{fig2} the evolution of both the high and low-mass binaries to
aid the comparison between the two classes of models.

\subsection{High-mass binaries}
\label{sec:bd_hm}

We start by considering the evolution of the high-mass binaries
\texttt{M1.62-B*}, some of which were already considered
in~\cite{Giacomazzo:2009mp}, where it was shown that initial magnetic
fields lower than $10^{14}\,\G$ do not affect the dynamics in the
inspiral phase. Overall, given the initial coordinate separation of $
45\,\km$, all binaries inspiral for approximately $3$ orbits before
merging at $t\approx 8.2\,{\rm ms}$. There are different ways to measure the time of the merger
and the one we adopt here consists in looking at the first peak in the
evolution of $|\Psi_4|$. This time corresponds approximately to when
the two stellar cores merge and we note that the external layers of
the stars enter into contact about $2\,{\rm ms}$ earlier. In the top
panel of the left column of Fig.~\ref{fig1} we show the evolution of
the maximum of the rest-mass density $\rho_{\rm max}$ normalized to
its initial value. It is particularly clear from the evolution of
$\rho_{\rm max}$ that all the models merge at the same time (e.g. see
the first minimum in the evolution), while the post-merger dynamics are
quite different. All the models form an HMNS that survives a few
milliseconds before collapsing to a Kerr BH, but its survival time
varies considerably, as well as the number of oscillations in the
evolution of the density before the rapid exponential increase in
correspondence with the collapse. A discussion about this will be
presented in Sec.~\ref{sec:delay}. 

The middle panel of the left column of Fig.~\ref{fig1} shows instead
the maximum of the absolute value of the divergence of the magnetic
field. To the best of our knowledge this is the first time that the
evolution of the divergence of the magnetic field is shown in a GRMHD
simulation of BNSs. Because this is a fundamental quantity to evaluate
the quality of a numerical calculation, we encourage other authors to
present it systematically as well. As expected on mathematical
grounds, the implementation of the GRMHD equations discussed in
Sect.~\ref{sec:Einsten_MHD_eqs} is such that the divergence of the
magnetic field is essentially at machine precision at all times. It is
important to stress that such a small violation would not be possible
with the cell-centered AMR algorithm provided by the \texttt{Carpet}
code unless the vector potential is used as an evolved variable.

\begin{figure*}
  \begin{center}
    \includegraphics[angle=0,width=7.0cm]{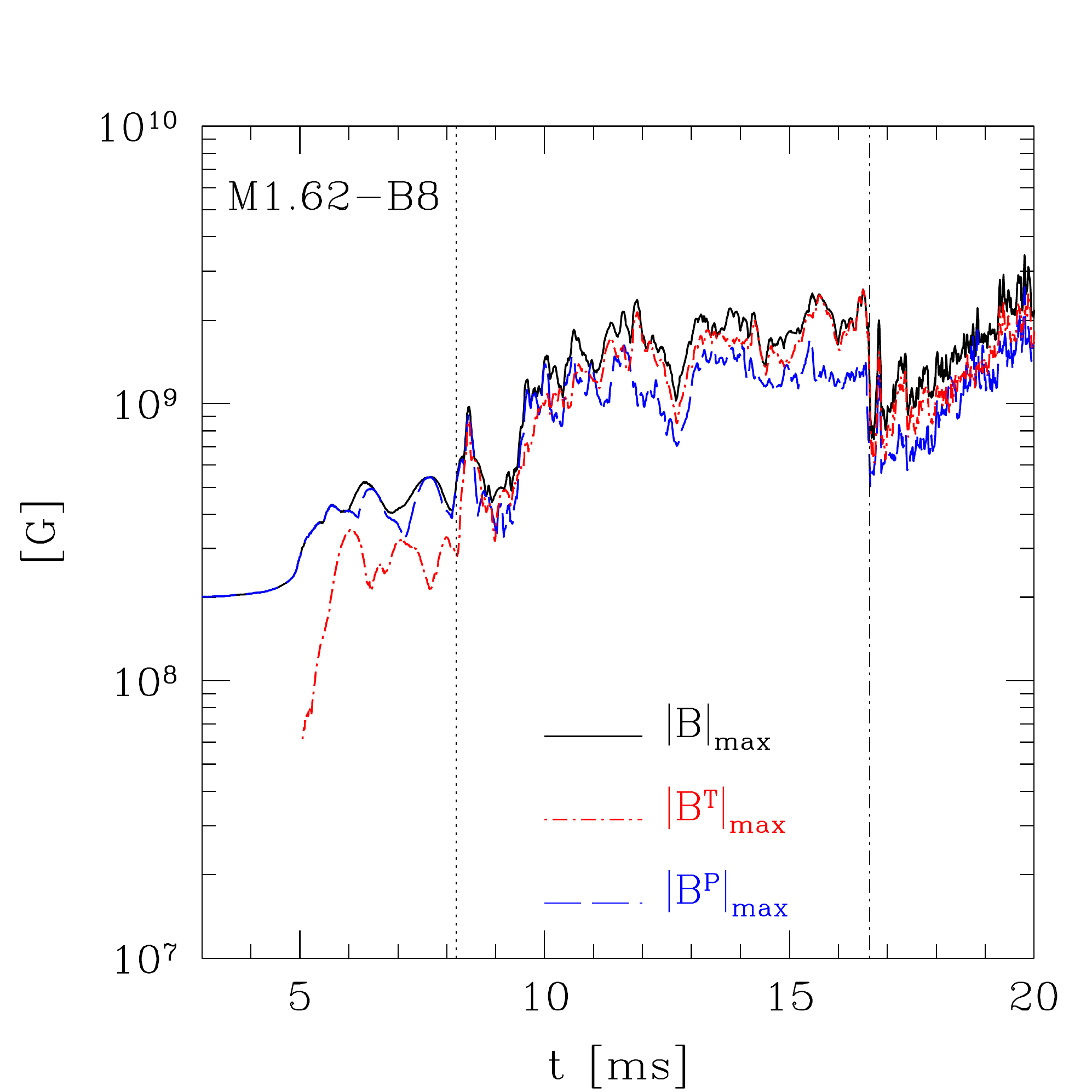}
    \includegraphics[angle=0,width=7.0cm]{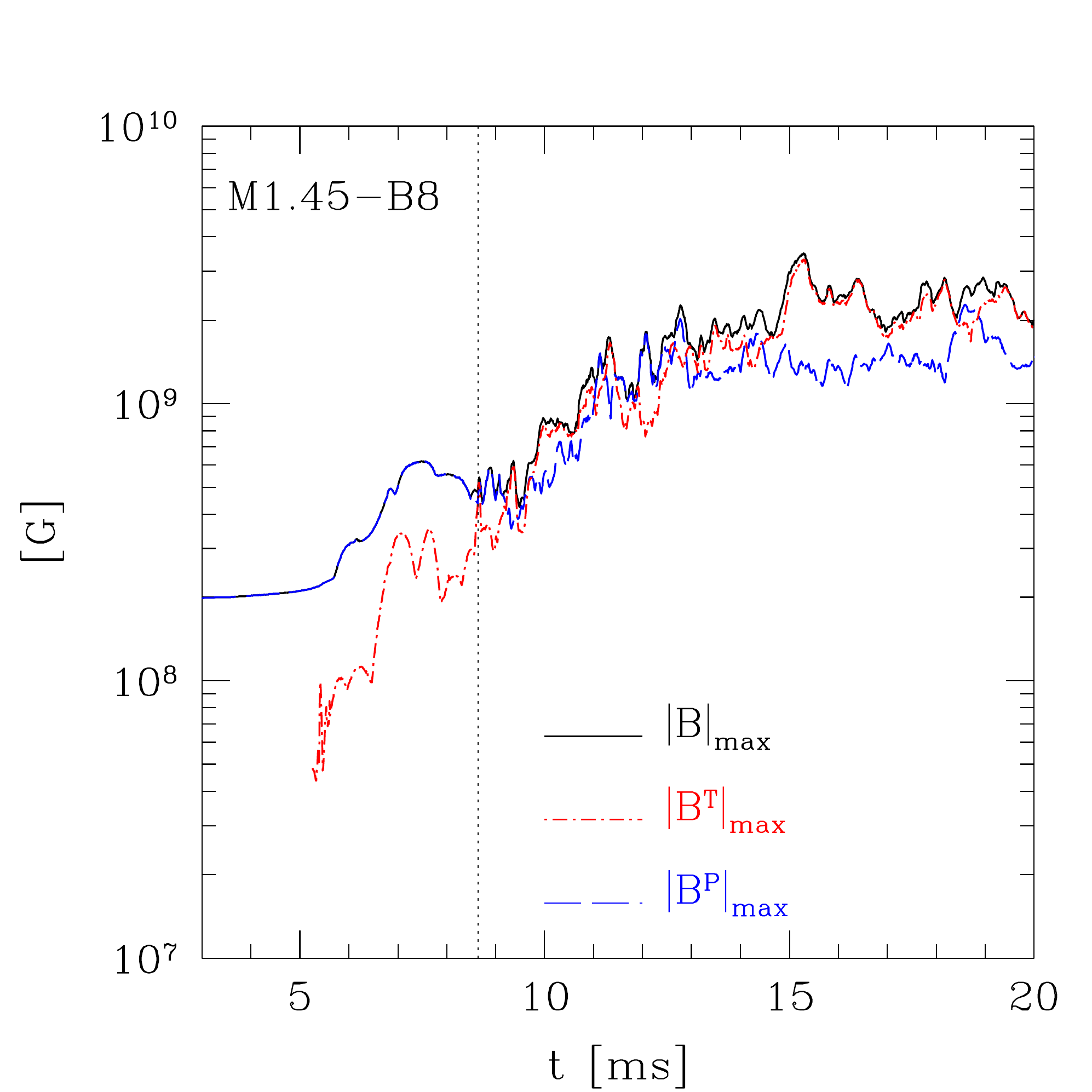}
    \includegraphics[angle=0,width=7.0cm]{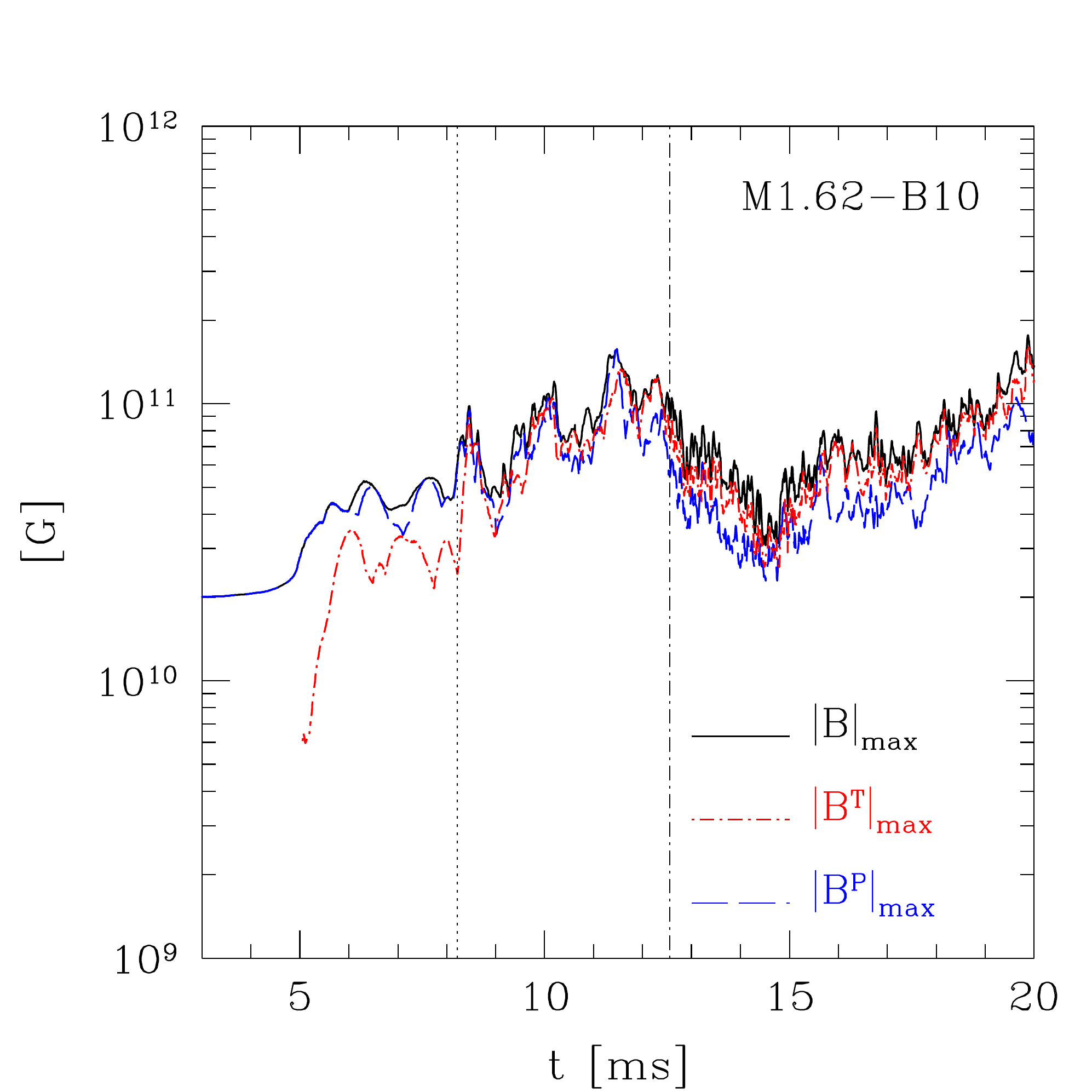}
    \includegraphics[angle=0,width=7.0cm]{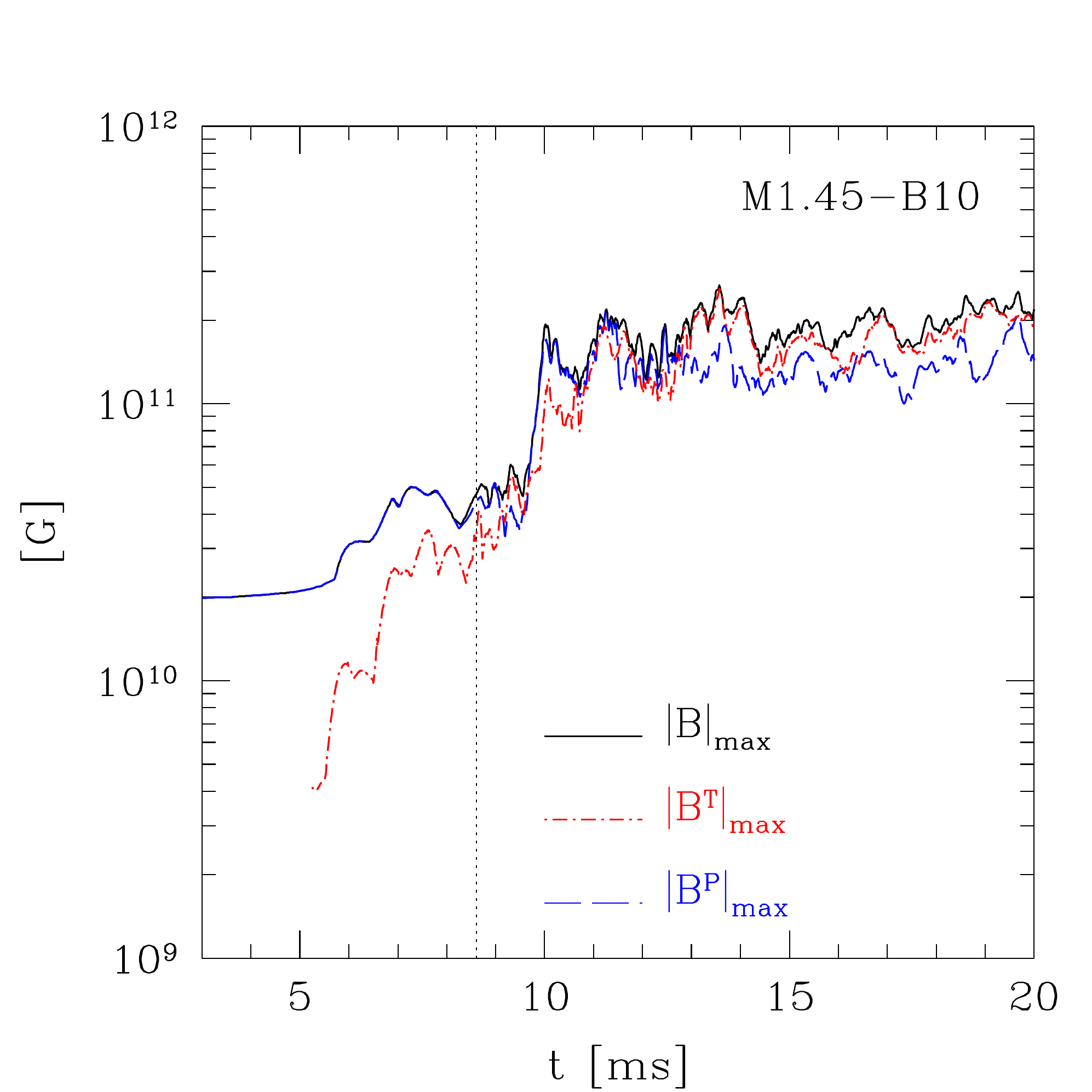}
    \includegraphics[angle=0,width=7.0cm]{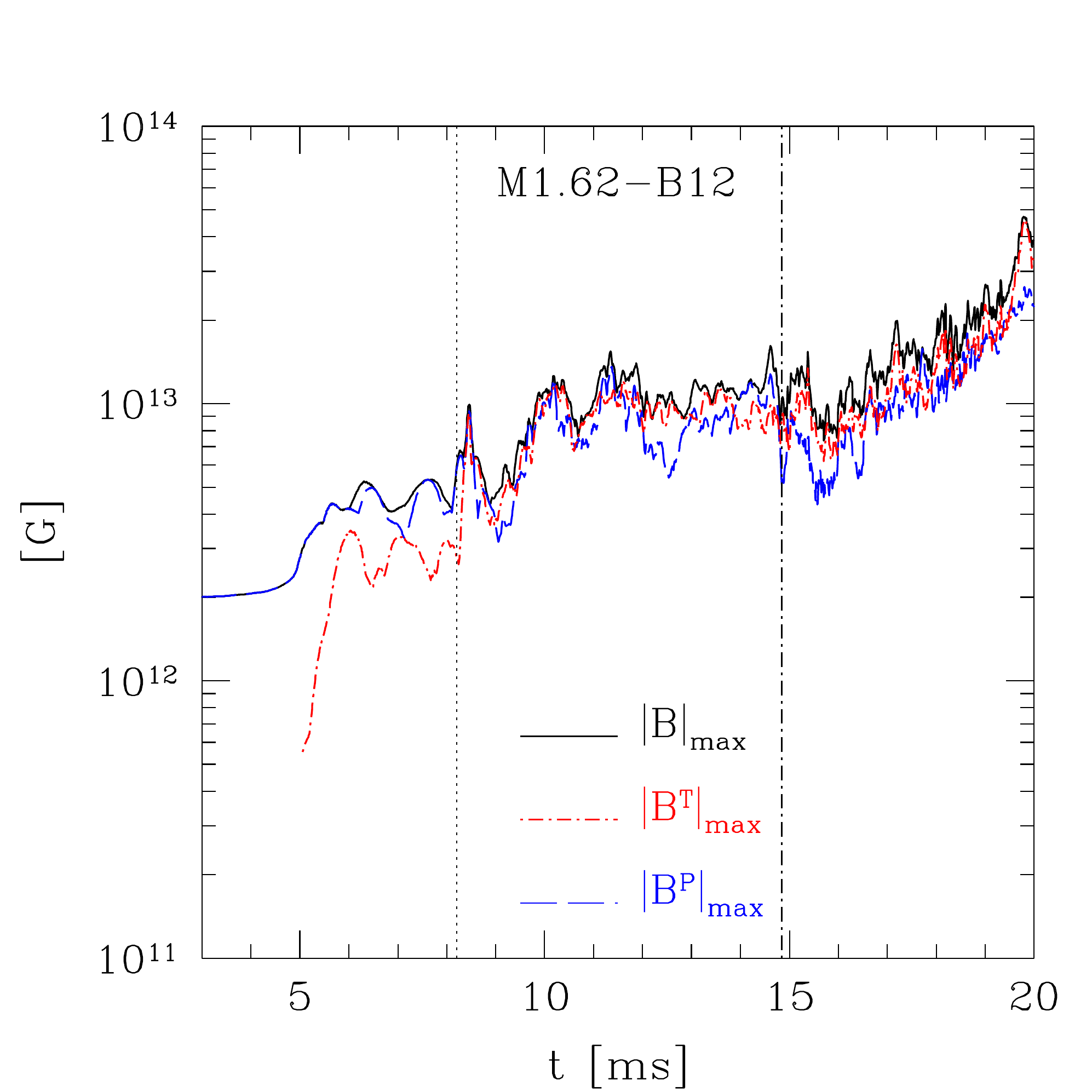}
    \includegraphics[angle=0,width=7.0cm]{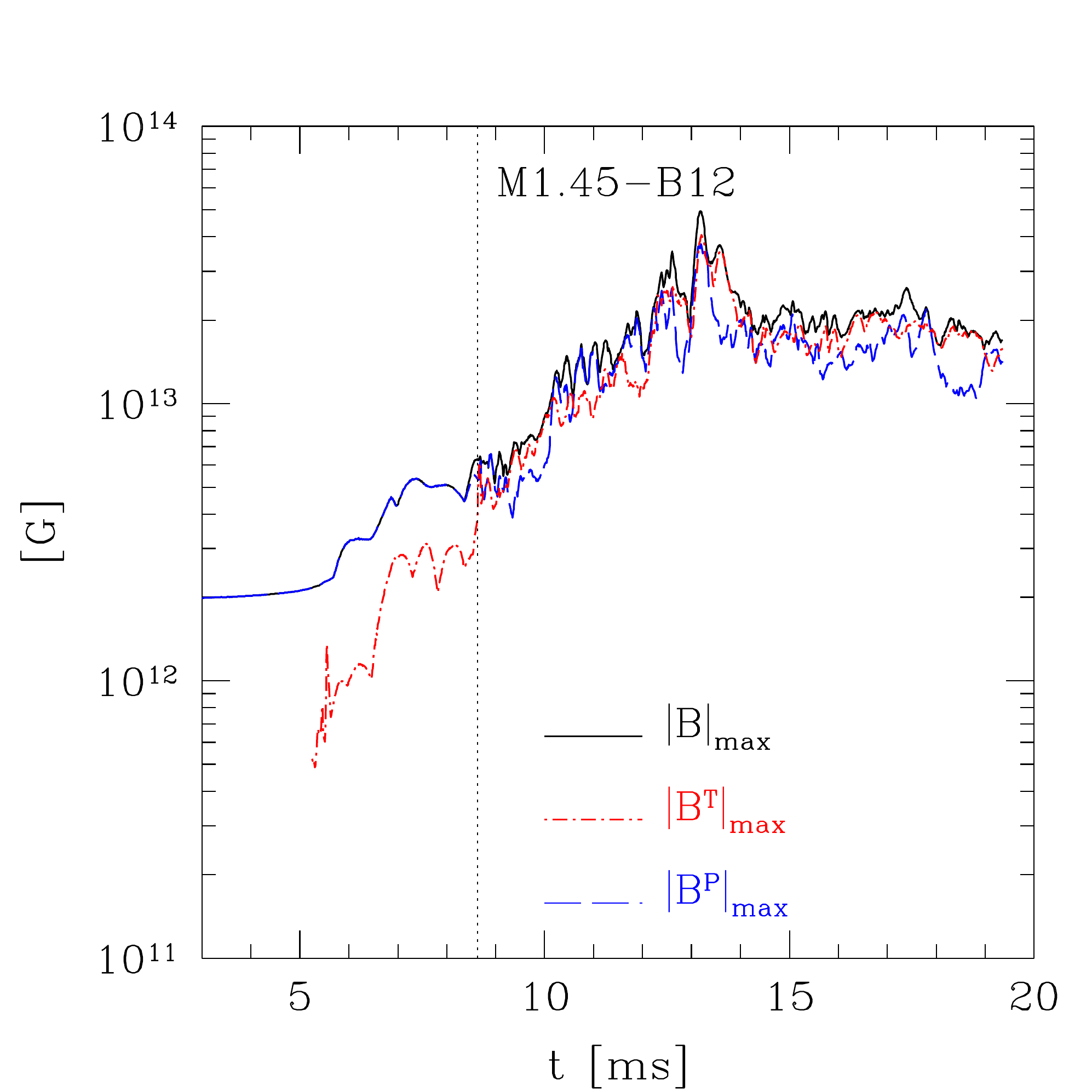}
  \end{center}
    \vskip -0.5 cm
  \caption{\label{fig2}Evolution of the maximum magnetic field
    strength $|B|_{\rm max}$ (black solid line) and of its poloidal
    $|B^{\rm P}|_{\rm max}$ (blue long-dashed line) and toroidal
    $|B^{\rm T}|$ (red short-dashed line) components during and after
    the merger. The left column refers to the high-mass model while the
    right one to the low-mass case. The vertical dashed lines refer to
    the time of the merger and of the collapse (measured respectively
    as the first and last peaks in the evolution of $|\Psi_4|$). Since
    the simulations of the low-mass binaries were not carried on until
    the collapse, only the time of the merger is shown in the panels
    in the right column.}
\end{figure*}

\begin{figure}
  \begin{center}
    \includegraphics[angle=0,width=8.0cm]{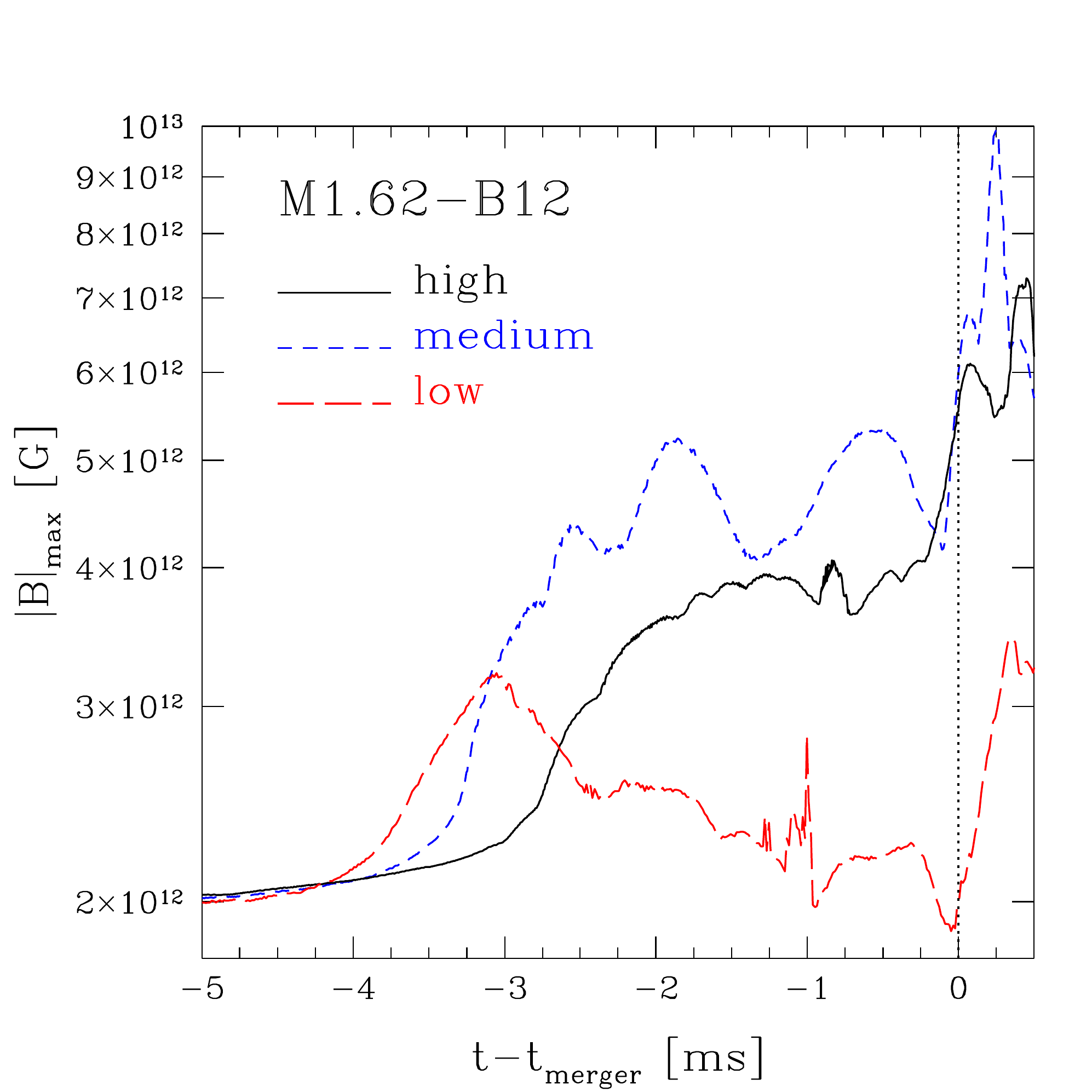}
  \end{center}
    \vskip -0.5 cm
  \caption{\label{fig2b}Evolution of the maximum magnetic field
    strength $|B|_{\rm max}$ for the high-mass model
    \texttt{M1.62-B12} evolved with three different resolutions:
    $h=177\,\mathrm{m}$ (high resolution, black solid line),
    $h=221\,\mathrm{m}$ (medium resolution, blue short-dashed line)
    and $h=354\,\mathrm{m}$ (low resolution, red long-dashed
    line). The curves have been shifted in time to account for the
    slightly different time of the merger.}
\end{figure}

Finally, the bottom panel of the left column of Fig.~\ref{fig1} shows
that the magnetic field grows mostly at the time of the merger and
reaches values which are about one order of magnitude higher, before
the collapse to BH. We note that in the case of the high-mass models
(left column) the values of $|B_{\rm max}|$ after BH formation refer,
for the large majority of the time, to matter outside the apparent
horizon and in the torus. This is because the steep gradients of the
matter variables inside the apparent horizon are under-resolved as a
result of the grid stretching and dissipated on a timescale which is
of the order of a fraction of a ms. Hence, with a few possible
exceptions, the data in the plots refers statistically to the matter
outside the apparent horizon.

The growth of the magnetic field at the merger is made more clear in
the different panels contained in the left column of Fig.~\ref{fig2},
where we concentrate in particular on the evolution of the maxima of
the total magnetic field (black solid line) and of its toroidal (red
dot-dashed line) and poloidal (blue long-dashed line) components. As
already shown in~\cite{Baiotti08}, Kelvin-Helmholtz instability
develops during the merger, when the external layers of the two NSs
enter into contact, \ie roughly $2\,{\rm ms}$ before the time of the
merger, which is indicated in those panels with the first vertical
dotted line. This purely hydrodynamical instability leads to the
formation of vortices that can curl magnetic field lines that were
initially purely poloidal and produce toroidal components. As it is
evident from the panels in Fig.~\ref{fig2}, a strong toroidal
component is indeed formed in all cases and it reaches values that are
comparable or larger than the poloidal component, but its energy is
not in equipartition with the kinetic energy in the layer. Despite the
exponential growth caused by the Kelvin-Helmholtz instability, the
overall amplification of the magnetic field is of an order of
magnitude at most, with a growth rate $dB/dt \simeq 2 \times 10^{12}\,
({\rm G/ms})$ in the case of model \texttt{M1.62-B12}. This is in
contrast with what was reported by~\cite{Price06}, where an amplification
of several orders of magnitude in the magnetic field of the HMNS was
observed, with a growth rate $dB/dt \simeq 2 \times 10^{15}\, ({\rm
  G/ms})$ for a model similar to \texttt{M1.62-B12}.

It is presently unclear what the origin of this discrepancy is. It is
possible that this is due to the use of very different numerical
techniques, namely smooth-particle hydrodynamics and HRSC methods. It
is also possible that although we have used the largest resolutions
employed so far in simulations of magnetized BNSs, such resolutions
are not yet sufficient to properly resolve the nonlinear development
of the instability. Studies of the effect of these instabilities and
of the consequent amplification of the magnetic fields have recently
been performed with local simulations on simpler
backgrounds~\cite{Zhang09,Obergaulinger10}. These studies have indeed
shown that in order to achieve convergence in the vortex region it is
necessary to use resolutions that are much higher than those
currently affordable in BNS simulations. On the other hand, by
performing simulations with different resolutions for model
\texttt{M1.62-B12} we did not observe any sensible difference in the
amplification of the magnetic field and indeed the magnetic field
evolution is certainly consistent if not convergent (see the
discussion in~\cite{Baiotti:2009gk} about why it is difficult to
determine the convergence order after the merger). This is shown in
Fig.~\ref{fig2b}, where we report the evolution of the maximum
magnetic field strength $|B|_{\rm max}$ for the high-mass model
\texttt{M1.62-B12} evolved with three different resolutions:
$h=177\,\mathrm{m}$ (high resolution, black solid line),
$h=221\,\mathrm{m}$ (medium resolution, which is the standard resolution used in
this article, blue short-dashed line) and
$h=354\,\mathrm{m}$ (low resolution, red long-dashed line). The curves
have been shifted in time to account for the slightly different time
of the merger. It is clear that doubling
the resolution produces a difference in the amplification of less than
a factor of about $2$ (compare the red long-dashed line with the black
solid line); the differences become even smaller when comparing the
medium and high resolution\footnote{Figure~\ref{fig2b} also shows a
  considerable increase in the magnetic field at the merger. However,
  this is not related to the Kelvin-Helmholtz instability, but rather
  to flux conservation which amplifies the magnetic field when the
  matter is compressed by the collision of the two stellar cores.}. A
similar consistency with resolution is not present in the simulations
reported in~\cite{Price06}, where the differences among the amplified
magnetic fields seem to become even larger with increasing
resolution\footnote{Note that because we are here capturing a
  non-sustained turbulent flow, the variations of the magnetic field
  strength with resolution are not necessarily monotonic.}.

Overall we believe that the main reason why the toroidal magnetic
field in our simulations does not grow significantly at the merger is
that the timescale over which the instability can develop is rather
short. The shear layer between the two stars, in fact, survives only
for about $1\,\ms$, before being destroyed by the collision between
the two stellar cores. In Refs.~\cite{Zhang09,Obergaulinger10}, the
amplification of the magnetic field has been observed on timescales
that are even shorter than this one, but only under very specific
conditions and only for specific values of the velocity at the shear
layer. The differences between the condition under which the
instability develops in our fully general-relativistic simulations and
those used in these local simulations may explain the different
results. Clearly, the best way to assess whether or not the
development of the Kelvin-Helmholtz instability leads to a large or
only to a moderate field amplification is to perform direct comparisons
with other general-relativistic simulations of magnetized
BNSs. Unfortunately, so far the only other reported evolution of the
magnetic field is the one in~\cite{Giacomazzo:2009mp}, which is
clearly not useful for an independent comparison.

As a final remark, it is important to emphasize that the toroidal and
poloidal components have comparable values also in the torus that is
formed after the collapse to BH (\cf panels in the left column of
Fig.~\ref{fig2}). Since most of the simulations to date of magnetized
accretion disks around BHs that model the central engine of short GRBs
use initial conditions in which the magnetic field has only a poloidal
component, it is of particular importance to remark that more
realistic initial data should instead have a toroidal and a poloidal
component of comparable magnitude.

\subsection{Delay of the Collapse}
\label{sec:delay}

It has been shown and discussed in a number of works that when the
merger leads to a collapse, the time of survival of the HMNS depends
on several factors, which include: the EOS, the efficiency in the
redistribution of angular momentum, and the efficiency of the radiative
transfer. Clearly, all of these influencing factors will act
differently in highly-magnetized matter and hence the delay time
$\tau_{\rm d}$, \ie the time between the formation of the HMNS and its
collapse to a BH, can be used to measure indirectly the magnetic
fields of the progenitor NSs. There are several different ways of
defining $\tau_{\rm d}$, but a convenient and gauge-invariant one is
to consider the delay time as the interval between the first and last
peak in the evolution of $|\Psi_4|$, which are always well-defined in
the amplitude evolution, as these can be taken to correspond to the
merger of the stellar cores and to the BH production.

\begin{figure}
  \begin{center}
   \includegraphics[angle=0,width=8.0cm]{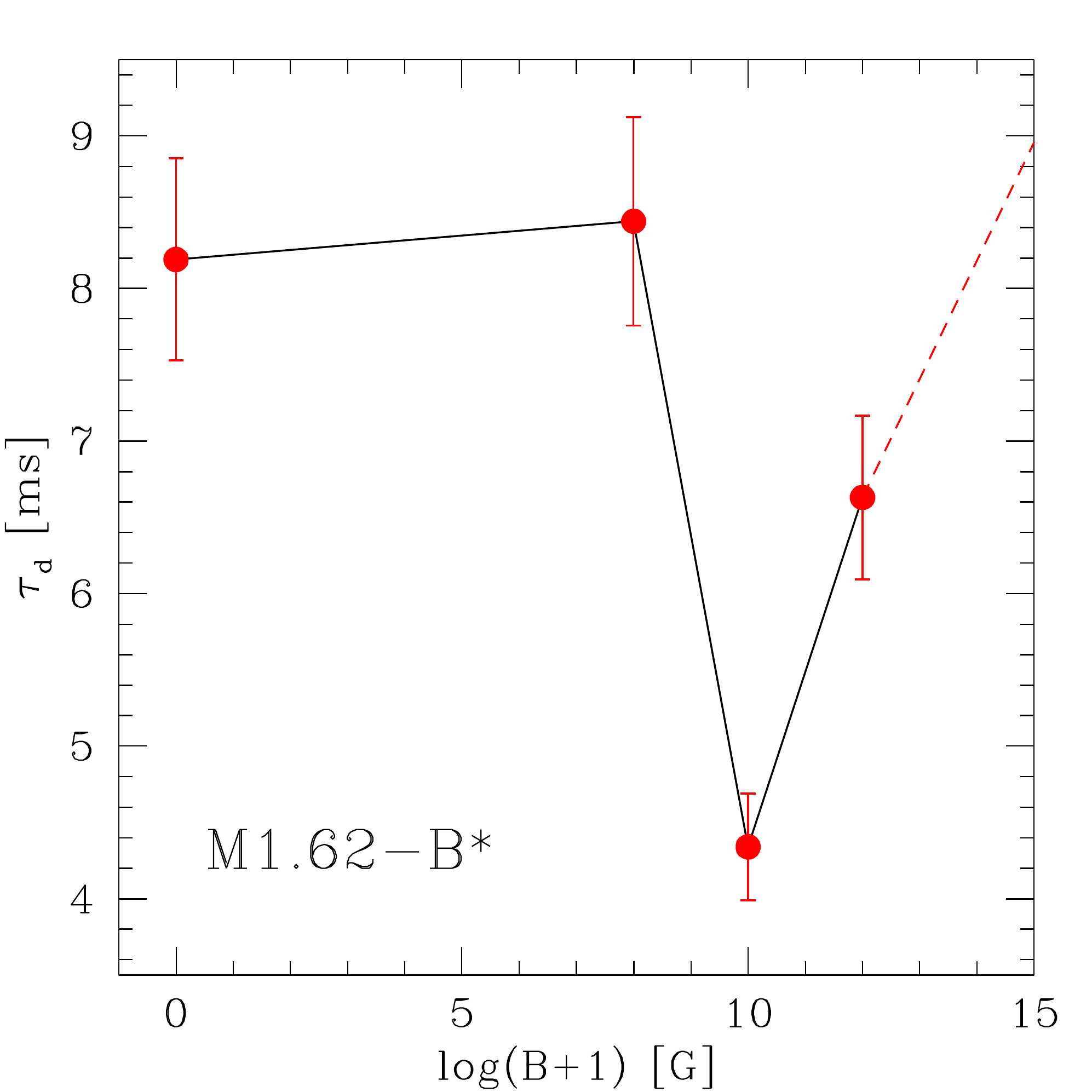} 
  \end{center}
    \vskip -0.5 cm
  \caption{\label{fig:delaytimes}Lifetime of the HMNS formed after the
    merger in the high-mass case as a function of the initial magnetic
    field. The error bar has been estimated from a set of simulations
    of unmagnetized binary NS mergers at three different resolutions;
    in particular, we have assumed that the magnetized runs have the
    same relative error on the delay time of the corresponding
    unmagnetized model. Indicated with a dashed line is the
    continuation of the delay times to ultra-high magnetic fields of
    $10^{17}\,\G$.}
\end{figure}

In Fig.~\ref{fig:delaytimes} we show therefore the survival time of
the HMNS as a function of the initial magnetic field strength,
together with the error bar as estimated from a set of simulations of
unmagnetized binary NS mergers at different resolutions (the delay
time converges at first order, increasing with resolution). It is
clear from Fig.~\ref{fig:delaytimes} that while models
\texttt{M1.62-B0} and \texttt{M1.62-B8} have roughly the same
post-merger dynamics and the same collapse time (see also the top left
panel of Fig.~\ref{fig1}), it is also clear that models
\texttt{M1.62-B10} and \texttt{M1.62-B12} collapse earlier than the
unmagnetized one. To understand why this is the case, we recall that
magnetic fields can affect the dynamics of the HMNS as first shown in
axisymmetric evolutions of an isolated differentially rotating
HMNS~\cite{Duez:2005cj,Stephens:2006cn}. In essence, magnetic fields
can, via magnetic tension\footnote{We recall that in Newtonian ideal
  MHD the Lorentz force appearing in the equation for the conservation
  of momentum is given by
\begin{equation}
\label{mag_ten_pres}
\frac{1}{4 \pi \rho}\left[(\nabla \times {\vec B}) \times {\vec B}\right] =
\frac{1}{4 \pi \rho}\left[
({\vec B \cdot \nabla}){\vec B}
- \nabla \left(\frac{B^2}{2}\right)
\right]\,,
\end{equation}
where in the right-hand side the first term is the ``magnetic
tension'' along the field lines and the second one is the (isotropic)
``magnetic pressure''.}, redistribute the angular momentum,
transporting it outwards and reducing the amount of differential
rotation that is essential in supporting the HMNS against
gravitational collapse (we recall that a HMNS has, by definition, a
mass which cannot be sustained by the star if rotating uniformly). The
ratio between the magnetic tension and the pressure gradients scales
like the ratio between the magnetic pressure and the gas pressure, and
this ratio increases (although remaining less than one) after the
merger because the magnetic fields are stronger and the HMNS is more
extended and has smaller pressure gradients. As a result, magnetic
fields can ``accelerate'' the collapse of these models, but only if they are
sufficiently strong so that the magnetic tension can be comparable to or
larger than the normal pressure gradients. Hence, the efficiency in
angular-momentum redistribution will be proportional to the intensity
of the (square of the) magnetic field and this explains why the delay
time is essentially unchanged for small magnetic fields, such as $B_0
\lesssim 10^{8}\,\G$. For larger values, however, the magnetic fields
can influence the dynamics of the HMNS and decrease $\tau_{\rm d}$ as
shown by models \texttt{M1.62-B10} and \texttt{M1.62-B12}.

Interestingly, the HMNS relative to the binary \texttt{M1.62-B12}
survives longer than the \texttt{M1.62-B10} one. This should not be
entirely surprising since a very large magnetic field will also
introduce a magnetic pressure [\cf~eq.~\eqref{mag_ten_pres}], which
will provide an additional pressure support and thus either compensate
or even dominate the angular-momentum redistribution. Indeed, when
simulating a binary with an initial magnetic field of $B_0 \simeq
10^{17}\,\G$ we have found that the delay time increases and is even
larger than the one obtained in the absence of a magnetic field. This is not shown
directly in Fig.~\ref{fig:delaytimes}, which has been restricted to
realistic values of the magnetic field, but we have indicated
with a dashed line the continuation of the delay times to ultra-high
magnetic fields. Clearly, because of this tight correlation between
the degree of magnetization of the NS matter and the delay of the time of 
the collapse, the measurement of the latter via a GW detection will allow
to infer the former.

\begin{figure}
  \begin{center}
   \includegraphics[angle=0,width=8.0cm]{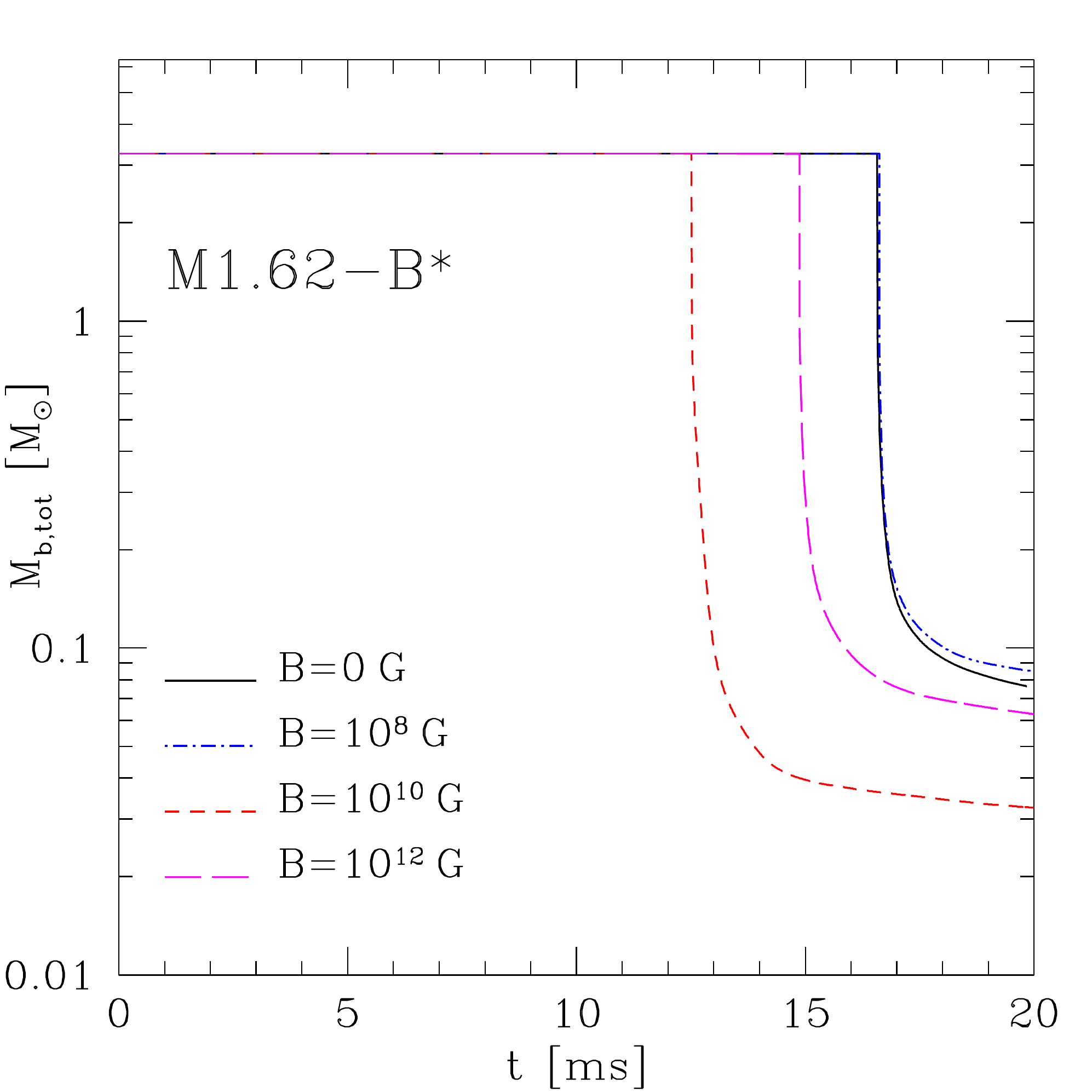} 
  \end{center}
    \vskip -0.5 cm
  \caption{\label{fig:totrestmass}Evolution of the total rest-mass for
    the different high-mass binaries considered. Note that the sudden
    drop corresponds to when the apparent horizon is formed since we
    exclude the region inside it from the computation of the
    mass. Note also that the early collapse of \texttt{M1.62-B10}
    leads to tori which are about a factor of two less massive.}
\end{figure}

\begin{figure}
  \begin{center}
   \includegraphics[angle=0,width=8.0cm]{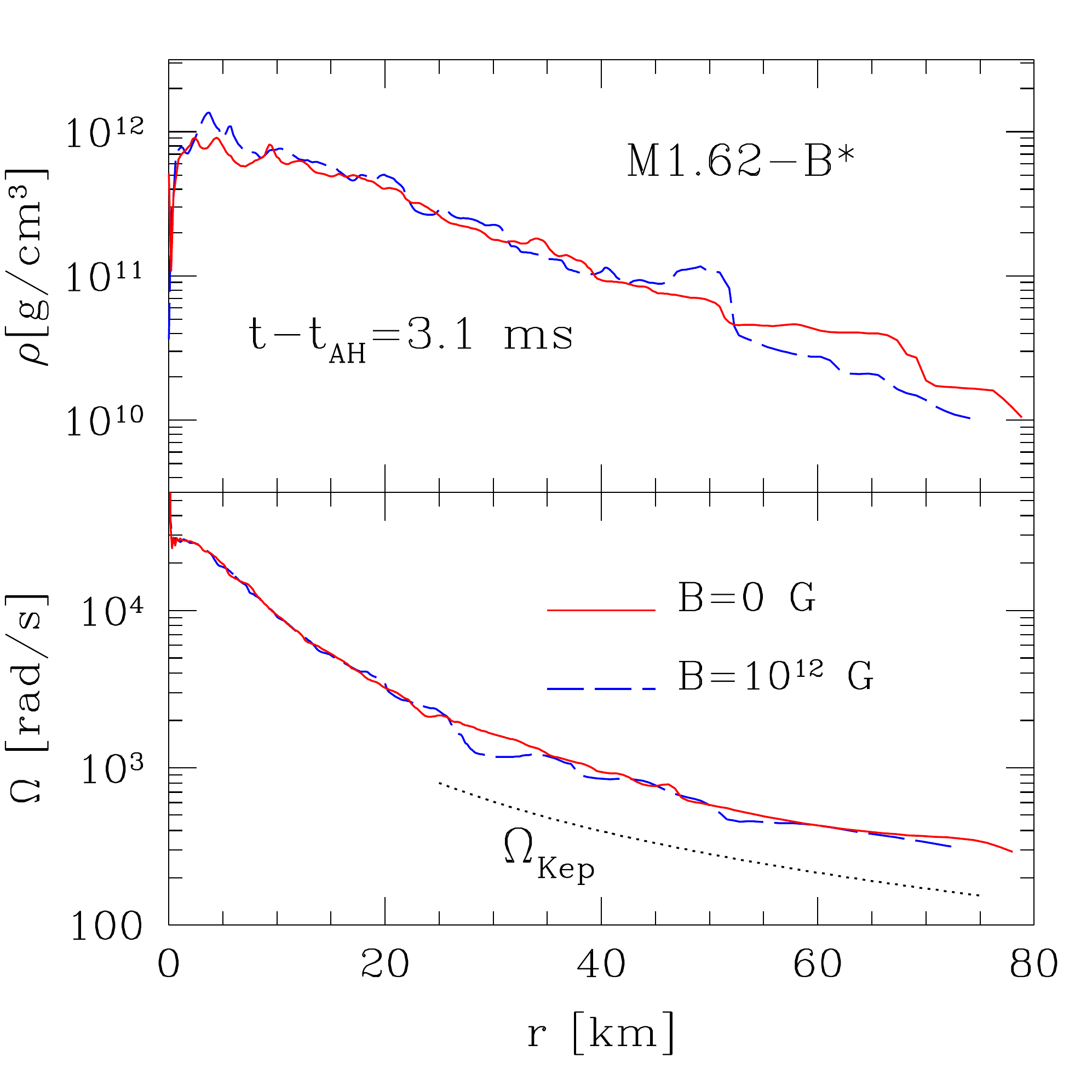} 
  \end{center}
    \vskip -0.5 cm
  \caption{\label{fig:rho_omega}Comparison of the properties of the
    tori produced either by a magnetized binary (\texttt{M1.62-B12},
    blue dashed line) or by a unmagnetized one (\texttt{M1.62-B0}, red
    solid line). \textit{Top panel:} rest-mass density along the
    $x$-axis at about $3\,{\rm ms}$ after the formation of the apparent
    horizon and which we truncate at $10^{10}\,{\rm
      g/cm}^3$. \textit{Bottom panel:} Angular velocity at the same
    time as above; shown as reference with a dotted line is the
    Keplerian angular velocity $\Omega_{\text{Kep}}$, which matches
    very well the outer parts of the torus.}
\end{figure}

The difference in the time of the collapse produces also small
differences in the mass of the final BH and torus. This is shown in
Fig.~\ref{fig:totrestmass}, which reports the evolution of the total
rest-mass for the different high-mass binaries considered, and where
the sudden drop corresponds to the formation of the apparent horizon
(the matter inside the horizon is excluded from the computation of the
baryon mass; see~\cite{Baiotti06} for a discussion of the properties
of the collapse with the gauge conditions used here). Similarly, in
Table~\ref{table:BH} we list the mass and spin of the BH formed at the
end of the evolution, and the mass and radius of the torus. Since the models
collapse at different times we have taken our measure at the end of
the simulation (\ie at $t\simeq 20\,\ms$), when the accretion onto the
BH is small and essentially stationary. In all cases the mass of the
BH is $M_{\rm BH}\approx 2.9 M_{\odot}$ and the spin is $a\equiv
J/M^2\approx 0.8$, but the mass of the torus drops from about
$0.063-0.085\,M_{\odot}$ to $0.033 M_{\odot}$ in the case of model
\texttt{M1.62-B10}. This is probably due to the fact that the magnetic
field causes some matter to move outside the core region and that will
become a BH; as a result, the longer the delay time, the larger the
tori.  We note that, even if small, these tori could still provide
sufficient energy to power short GRBs.

As a final remark we note that at least over the timescales considered
here, the differences in the local dynamics of the torus matter
between magnetized and unmagnetized binaries is very small. This is
because the magnetic field is not yet strong enough to produce
significant changes in the dynamics. A convincing example is shown in
Fig.~\ref{fig:rho_omega}, which offers a comparison of the properties
of the tori produced either by a magnetized binary
(\texttt{M1.62-B12}, blue dashed line) or by a unmagnetized one
(\texttt{M1.62-B0}, red solid line). The top panel, in particular,
shows the rest-mass density along the $x$-axis at about $3\,{\rm ms}$ after
the formation of the apparent horizon and which we truncate at
$10^{10}\,{\rm g/cm}^3$. Besides small differences (the data refers to
very different simulations), the density profiles are very similar. An
analogous conclusion can be drawn when looking at the bottom panel,
which shows the angular velocity at the same time as above; also
reported as reference with a dotted line is the Keplerian angular
velocity $\Omega_{\text{Kep}}$, which matches very well the outer
parts of the torus.  

\begin{table}
  \caption{\label{table:BH}Columns $2-3$ report the mass $M$ and spin
    $a$ of the BH, while column $4$ shows the mass of the torus formed
    after the merger of the high-mass models, and column $5$ the
    radius of the torus (computed as a mean of the position where the
    rest-mass density goes below $10^{10} \rm{gcm^{-3}}$ in the time
    interval between $19$ and $20$ ms). All the other quantities have
    been measured at $t=20\,\ms$, when the accretion onto the BH is
    small and essentially stationary.}
\begin{ruledtabular}
\begin{tabular}{lcccc}
Binary &
\multicolumn{1}{c}{$M~[M_{\odot}]$}&
\multicolumn{1}{c}{$a\equiv J/M^2$}&
\multicolumn{1}{c}{$M_{\rm tor}~[M_{\odot}]$}&
\multicolumn{1}{c}{$r_{\rm tor}~[\rm{km}]$}
\\
\hline
\texttt{M1.62-B0}   & $2.90$ & $0.80$ & $0.076$ & $105\pm 13$ \\
\texttt{M1.62-B8}   & $2.89$ & $0.80$ & $0.085$ & $102\pm 16$ \\
\texttt{M1.62-B10}  & $2.94$ & $0.82$ & $0.033$ & $ 69\pm  4$ \\
\texttt{M1.62-B12}  & $2.91$ & $0.81$ & $0.063$ & $ 94\pm  4$ \\
\end{tabular}
\end{ruledtabular}
\vskip -0.25cm
\end{table}

\subsection{Low-mass binaries}
\label{sec:poly_hm}

As already shown in~\cite{Giacomazzo:2009mp} and anticipated in the
previous Section, also in the low-mass case the presence of an initial
magnetic field introduces no significant modification in the evolution
of the binaries during the inspiral. To compare directly with the
behavior of the high-mass binaries, we show in the right column of
Fig.~\ref{fig1} the evolution of the maximum of the rest-mass density
normalized to its initial value, the maximum of the absolute value of
the divergence of the magnetic field and the maximum of the magnetic
field. We recall that in Ref.~\cite{Rezzolla:2010} it was shown that
the low-mass models take more than $100\,{\rm ms}$ to collapse to BH
in the unmagnetized case, so the $20$ ms of evolution of the present
work are not sufficient to reach the collapse.

In the top panel it is possible to appreciate that the evolution of
$\rho_{\rm max}$ for the unmagnetized case and those of the magnetized
binaries are very similar, with only small differences in the
frequency of the oscillations of the HMNS formed after the
merger. Overall, the presence of a magnetic field decreases the
oscillation frequency (\cf inset), probably because the additional
magnetic tension counters the expansions of the bar-deformed
HMNS. Moving over to the right bottom panel of Fig.~\ref{fig1}, it is
possible to note that also in the low-mass case all the magnetized
models show an amplification of the magnetic field of about one order
of magnitude and also in this case the divergence of the magnetic
field is zero essentially at machine precision. As for the high-mass
binaries, interesting point to note is that, on the timescale studied
here, the magnetic field grows by about one order of magnitude soon
after the merger (at $t\approx 8 {\, \rm ms}$), but then it saturates
to a constant value.

\begin{figure*}
\begin{center}
   \includegraphics[angle=0,width=8.0cm]{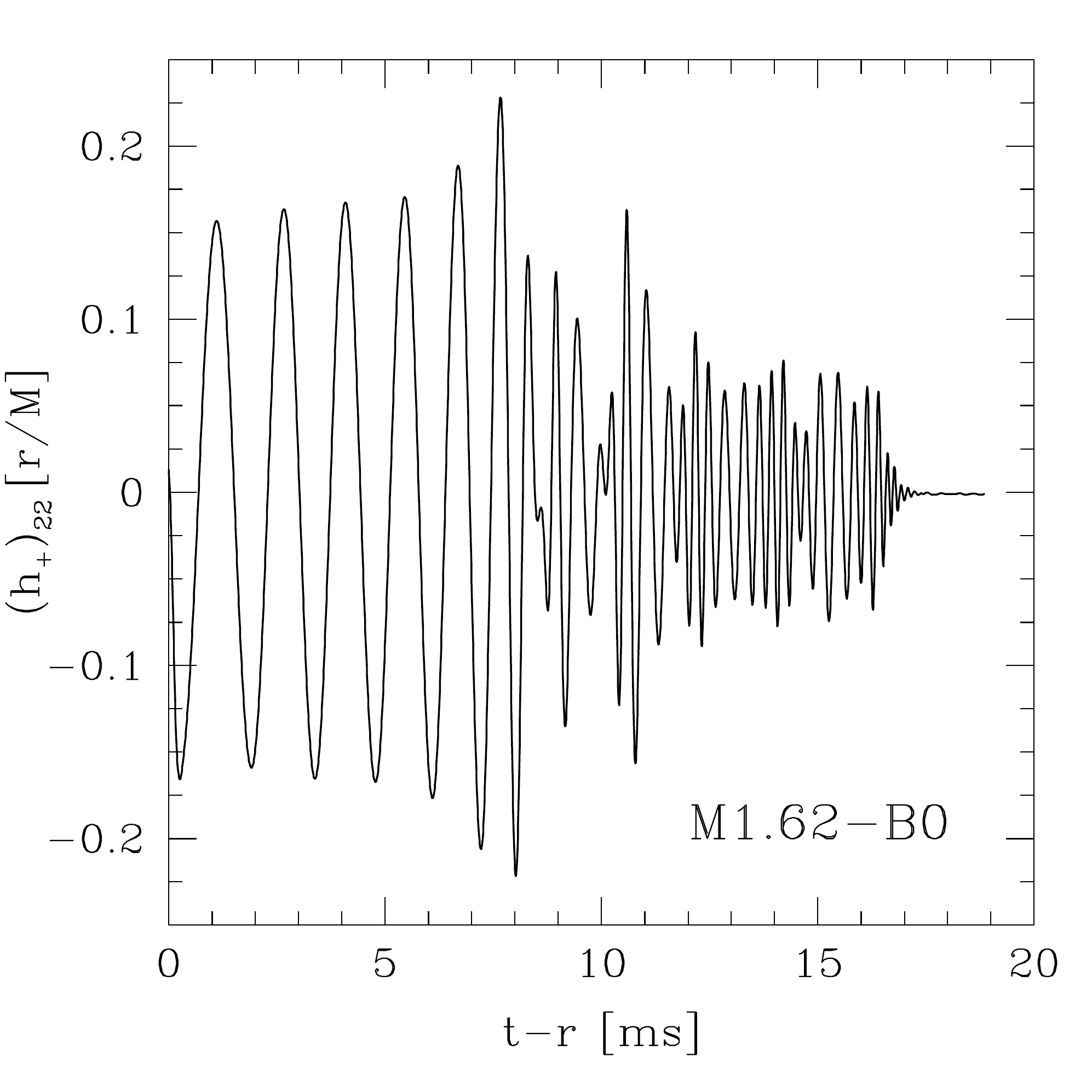} 
   \includegraphics[angle=0,width=8.0cm]{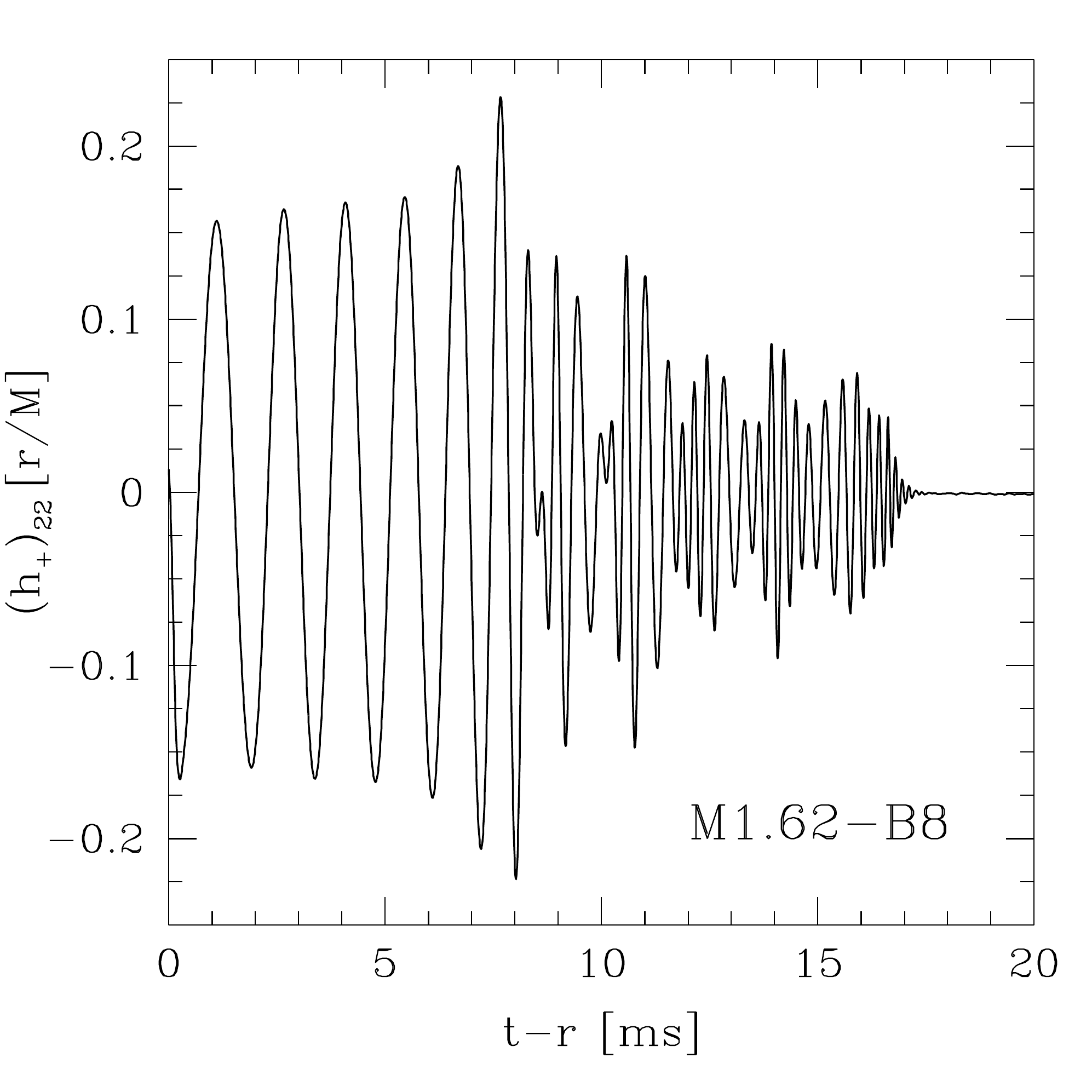}
   \includegraphics[angle=0,width=8.0cm]{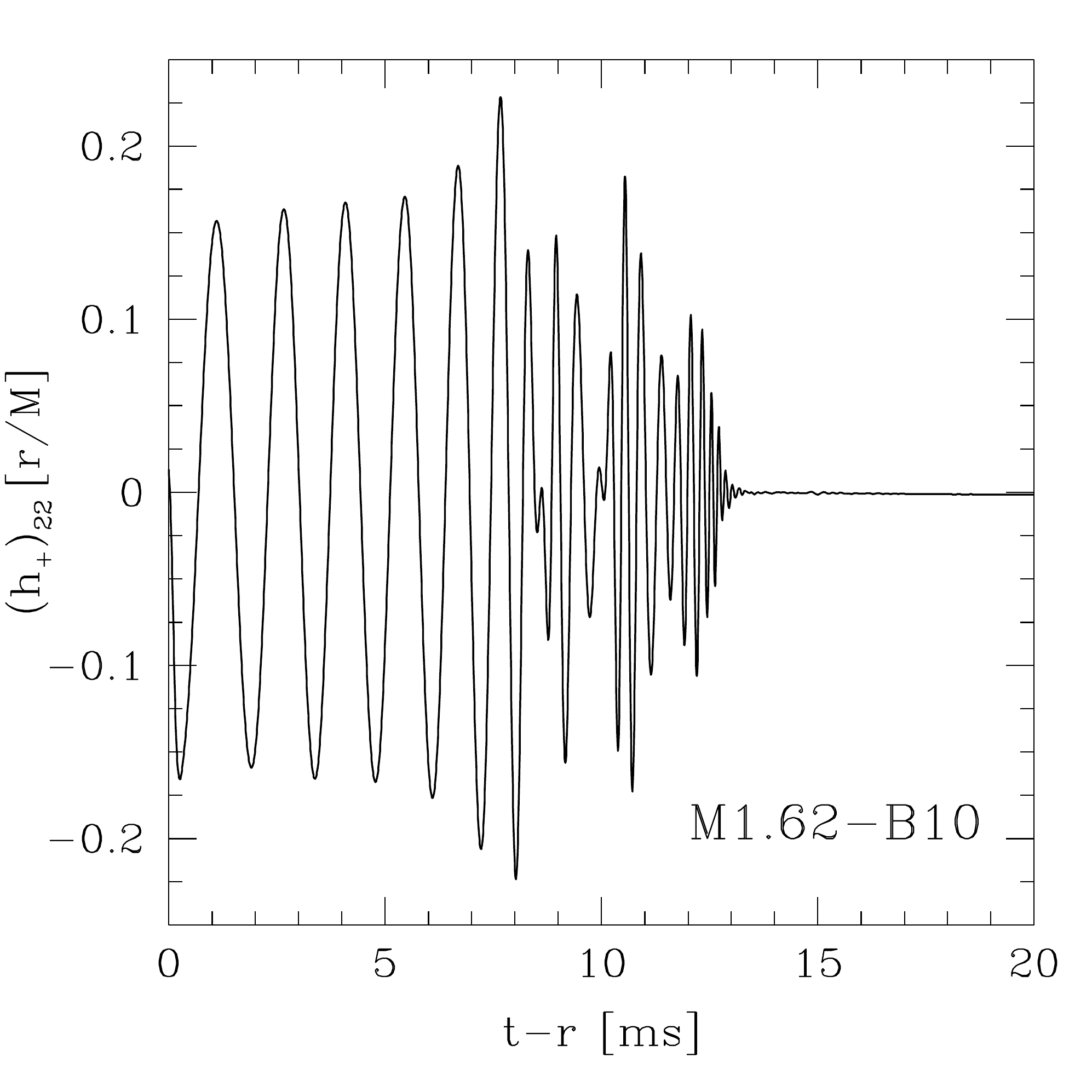} 
   \includegraphics[angle=0,width=8.0cm]{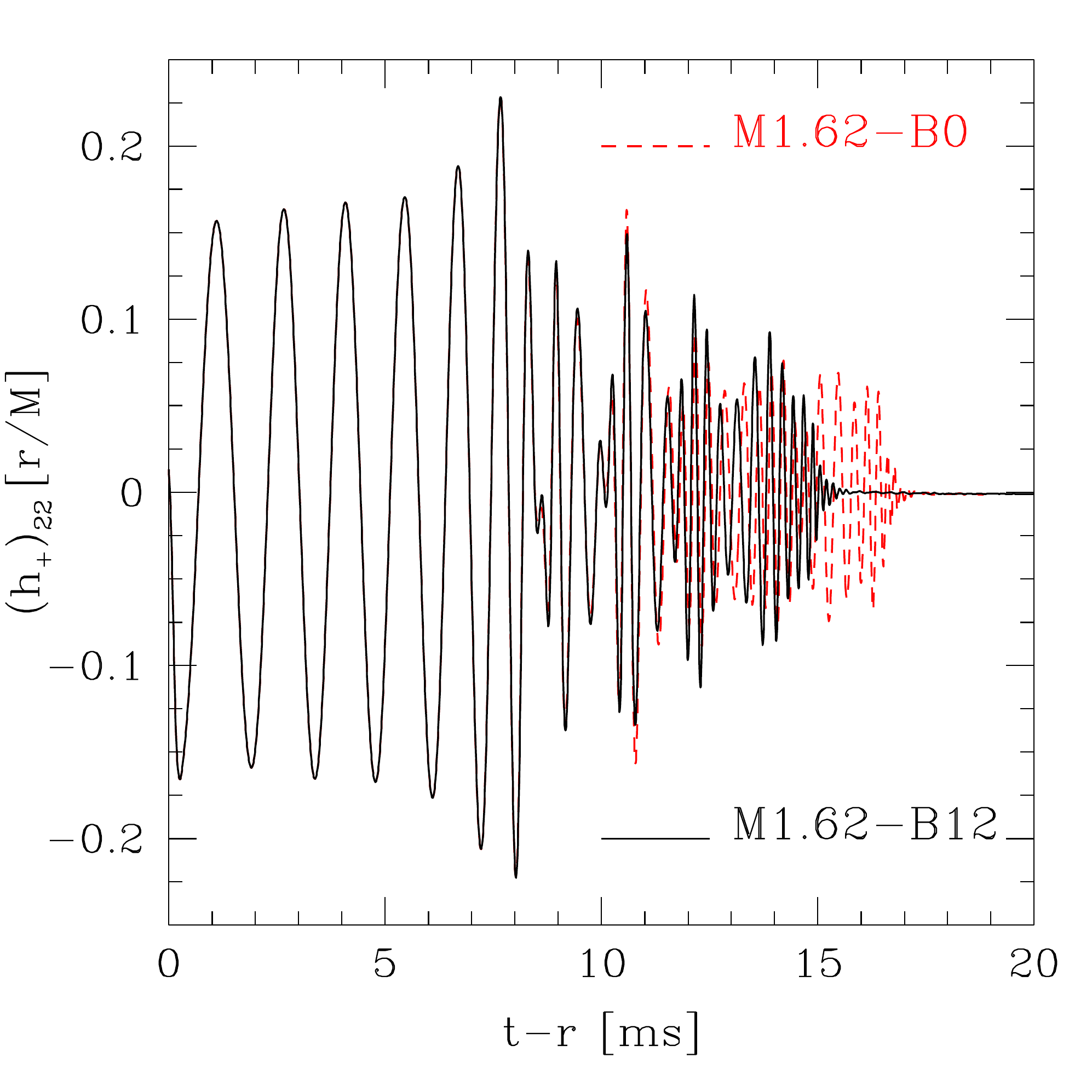} 
\end{center}
    \vskip -0.5 cm
   \caption{Gravitational waves for the high-mass binaries as a
     function of the retarded time $t-r$ in ms. The last panel shows
     for comparison also the unmagnetized model (\ie red dashed line
     which collapses at $t-r\approx 17\,\ms$) together with the model
     \texttt{M1.62-B12} (black solid line which collapses
     earlier).\label{fig:h22_HM}}
\end{figure*}

\begin{figure*}
\begin{center}
   \includegraphics[angle=0,width=8.0cm]{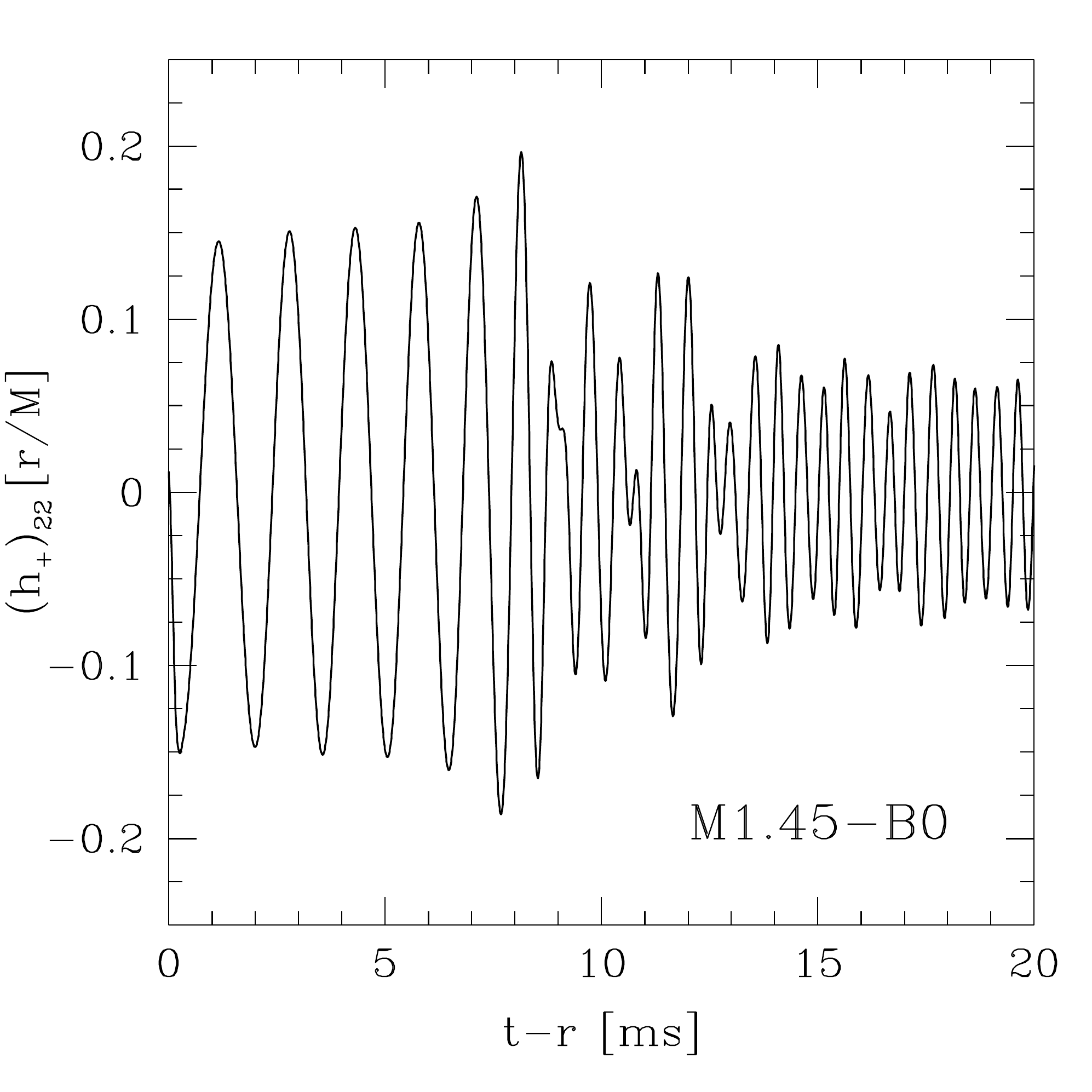} 
   \includegraphics[angle=0,width=8.0cm]{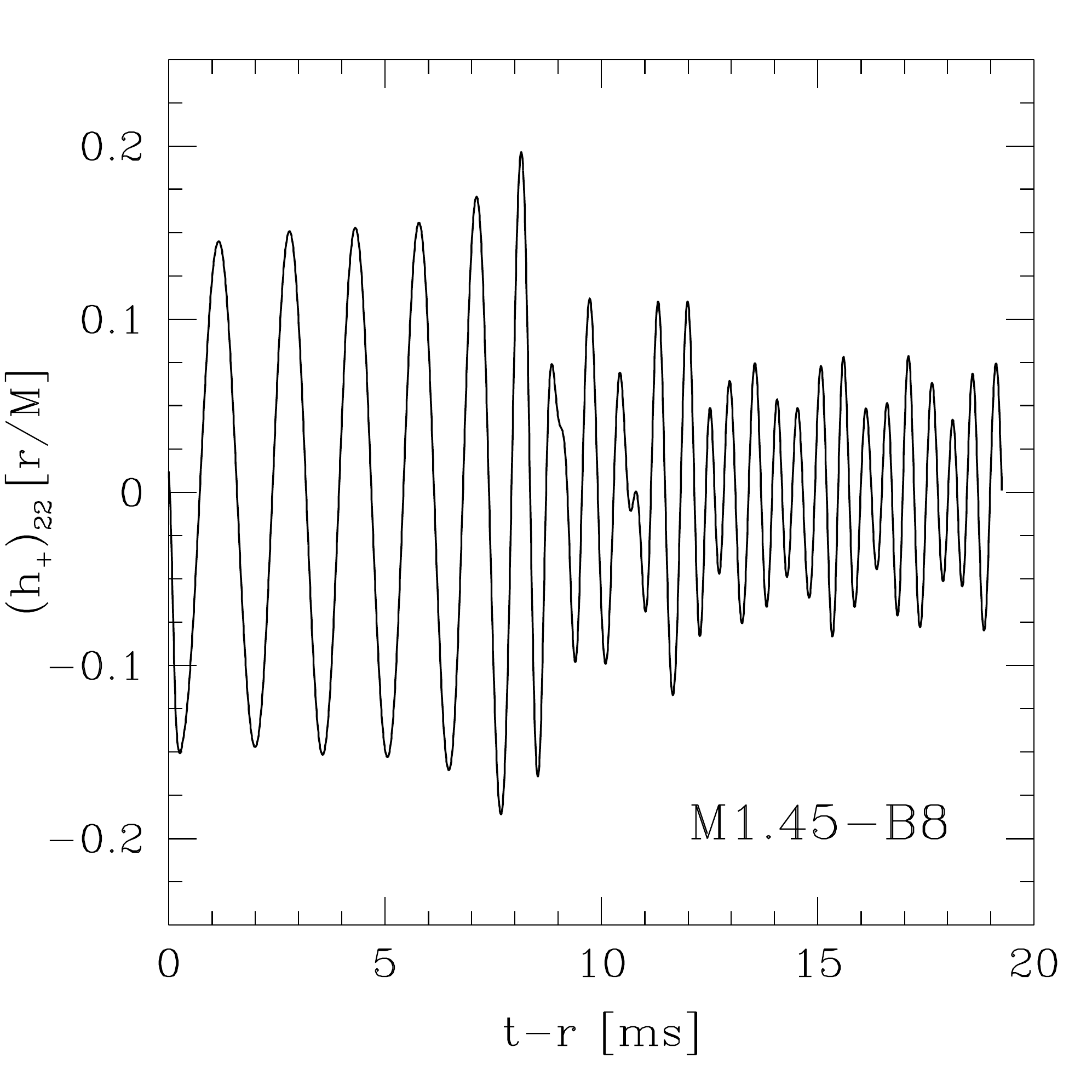}
   \includegraphics[angle=0,width=8.0cm]{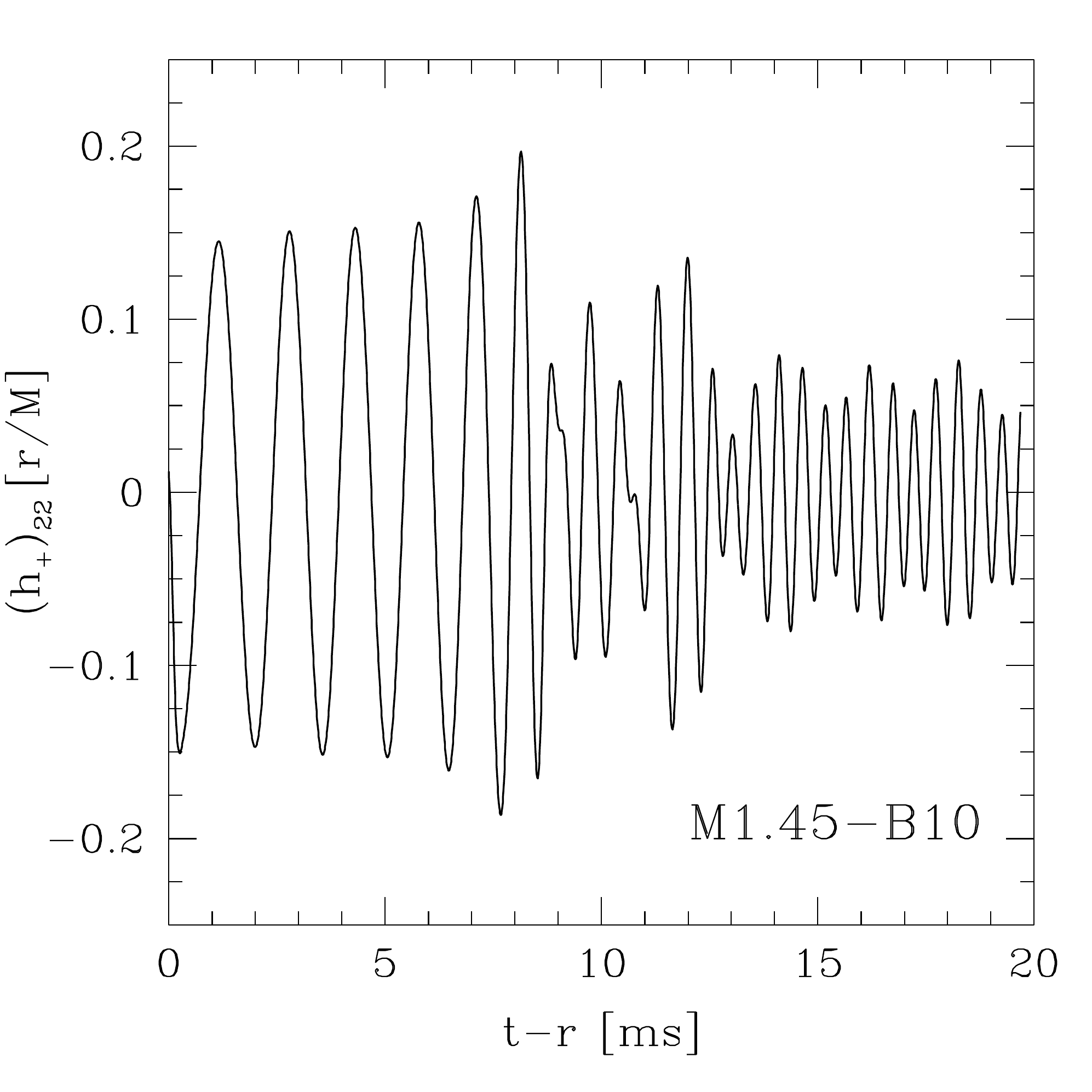} 
   \includegraphics[angle=0,width=8.0cm]{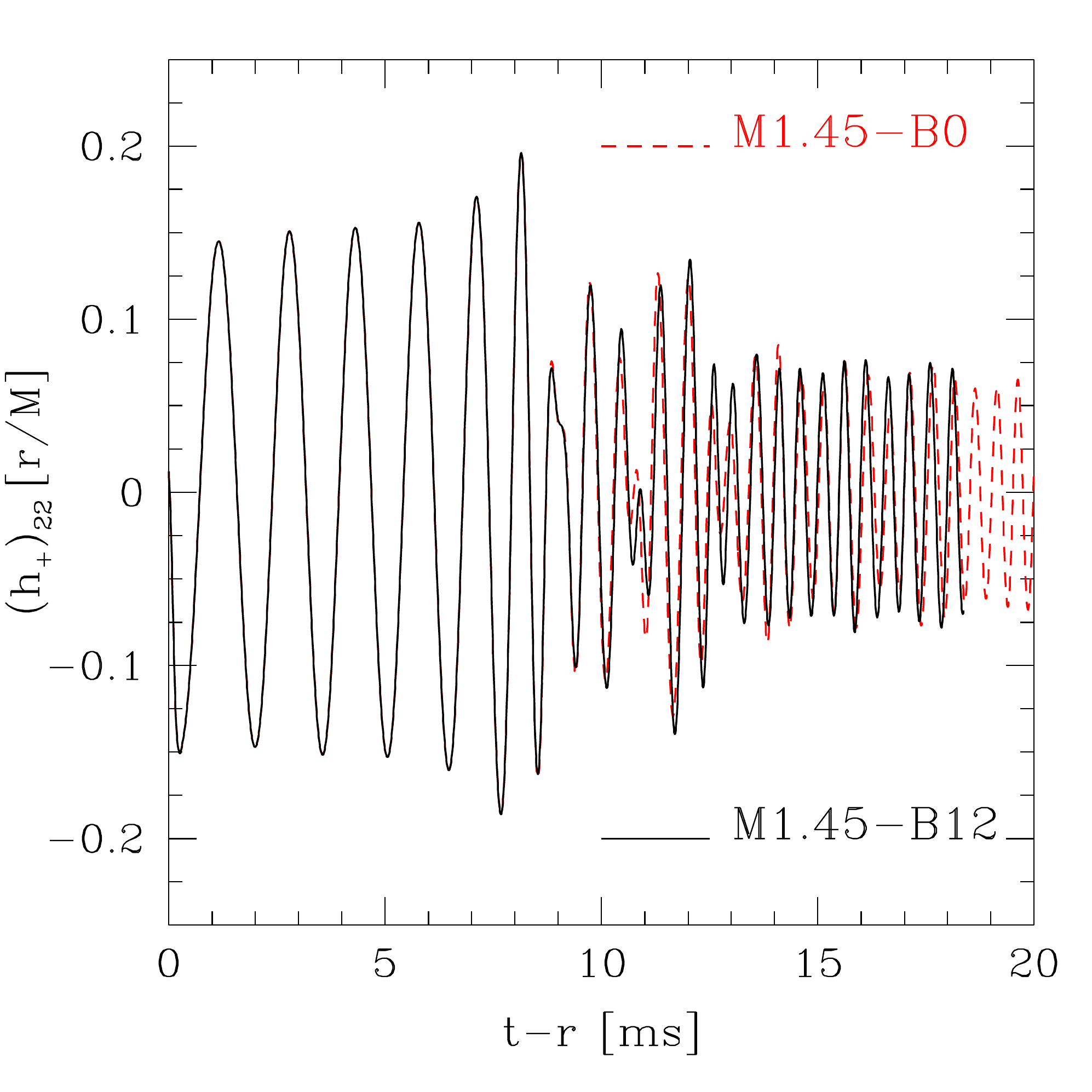} 
\end{center}
    \vskip -0.5 cm
   \caption{Gravitational waves for the low-mass binaries as a
     function of the retarded time $t-r$ in ms. The last panel shows
     for comparison also the unmagnetized model (\ie red dashed line which
     terminates at $t-r=20\,\ms$) together with the model
     \texttt{1.45-B12} (black solid line which terminates
     earlier).\label{fig:h22_LM}}
\end{figure*}

Additional information about the magnetic-field evolutions for the
three different models are given in the panels in the right column of
Fig.~\ref{fig2}. Also in this case the toroidal component of the
magnetic field (red dot-dashed line) is amplified exponentially
because of the Kelvin-Helmholtz instability at the time of the merger
of the external layers of the stars and it reaches the same value of
the poloidal component. Both components have comparable values for the
remaining duration of the simulation and we expect that also in this
case the collapse of the HMNS will produce a torus with a magnetic
field configuration in which the toroidal and poloidal components have
the same strength. This seems to be, at least for the equal-mass BNSs
considered here, a universal characteristic of the tori that are
formed from these systems.

\section{Gravitational-wave emission}
\label{sec:gw}

\subsection{High-mass binaries}
\label{sec:gw_hm}

In Fig.~\ref{fig:h22_HM} we show the GW signals emitted by the $4$
high-mass binaries considered in this paper. The top left panel shows
the unmagnetized case, the top right panel the model with an initial
magnetic field of $10^{8}\,\G$, the bottom left panel
$B\approx10^{10}\,\G$ and the bottom right panel
$B\approx10^{12}\,\G$. In the bottom right panel, together with
\texttt{M1.62-B12} (black solid line, which collapses at $t\approx
16\,{\rm ms}$), we also show - to make the comparison clearer - the evolution
of~\texttt{M1.62-B0} (red dashed line, which collapses later). All the
waveforms exhibit very similar features and, with the exception for
the different duration of the post-merger phase already discussed in
Sec.~\ref{sec:bd_hm}, they are almost indistinguishable from each
other. Therefore, for all the models the signal is essentially
composed of three parts: the inspiral (from $t-r=0\,{\rm ms}$ to
$t-r\approx 8\,{\rm ms}$), the HMNS evolution (from $t-r\approx
8\,{\rm ms}$ to $t-r\approx 13-17\,{\rm ms}$) and the ring-down of the
final BH. The high-frequency oscillations in the post-merger phase are
due to the cores of the two NSs that repeatedly bounce against each
other until a sufficient amount of angular momentum is extracted via
GWs emission or is moved to the external layers of the HMNS via the
magnetic-field tension. When this happens, the centrifugal support becomes
insufficient to balance the gravitational forces and the HMNS is induced
to collapse to a rotating BH with dimensionless spin $J/M^2 \simeq
0.80$ (\cf Table~\ref{table:BH}). Such oscillations are directly
related to the oscillations visible in the evolution of the maximum of
the rest-mass density in the top-left panel of Fig.~\ref{fig1}.

\subsection{Low-mass binaries}
\label{sec:gw_lm}

In analogy with what was done for the high-mass binaries, we show in
Fig.~\ref{fig:h22_LM} the GW signal for the low-mass models and also
in this case the bottom right panel shows both the \texttt{M1.45-B12}
(black solid line terminated at $t-r\approx 19 {\, \rm ms}$) and
\texttt{M1.45-B0} (red dashed line) models for comparison. Since we
have not evolved these models until the collapse of the HMNS to BH,
only the inspiral and the post-merger phase (the part of the signal
for $t-r\gtrsim 8.5\,{\rm ms}$) are present in the GW signal. The
high-frequency oscillations in the post-merger phase are related to
the formation of a bar-deformed HMNS (as already described
in~\cite{Baiotti08}), whose spinning frequency is not significantly
affected by the presence of magnetic fields. Also in this case, all
the waveforms are very similar to each other both during the inspiral
and after the merger. As a result, and in contrast with what was seen for
the high-mass case, the differences in the phase evolution are very
small, at least over the timescales considered here (\cf bottom right
panel). Clearly, if the HMNS continues to exist for longer times (on
the radiation-reaction timescale), then the small differences may grow
sufficiently and lead to a detectable difference. While the numerical
simulation of the secular evolution of the HMNS represents a challenge
that we will address in future work, its impact on the detectability
of the magnetic field will be further discussed in the next Section.

\begin{figure*}
\begin{center}
   \includegraphics[angle=0,width=7.0cm]{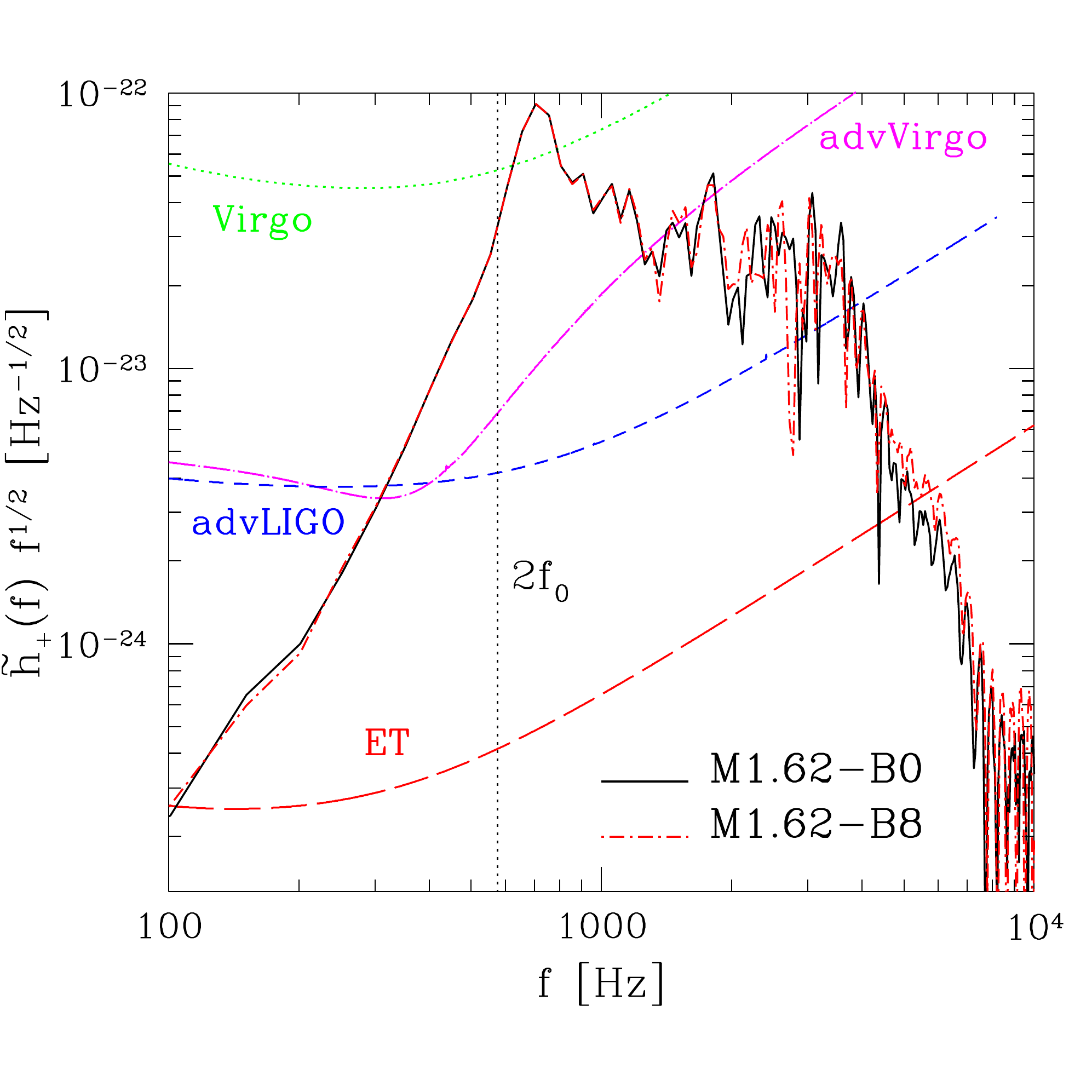} 
   \includegraphics[angle=0,width=7.0cm]{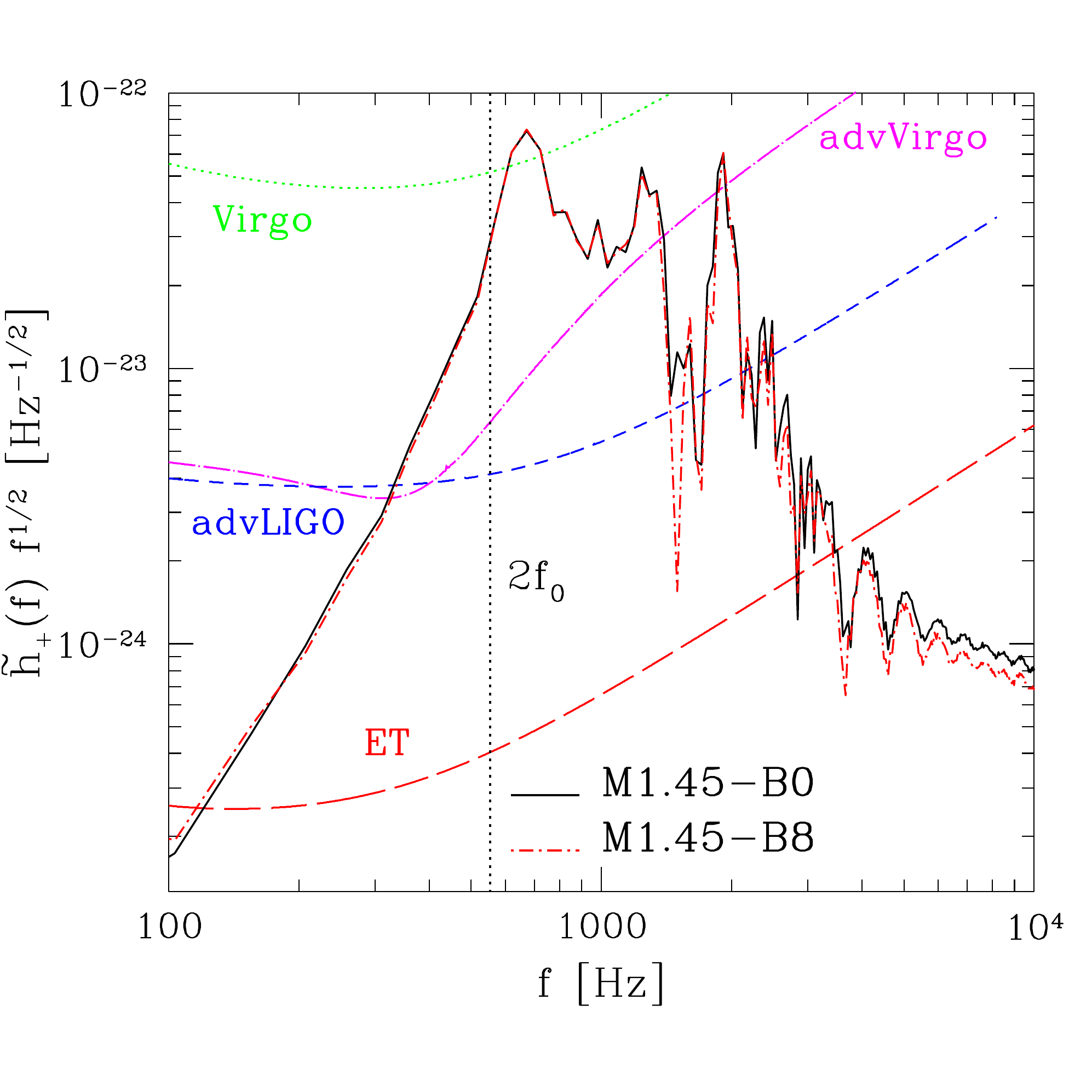}
   \includegraphics[angle=0,width=7.0cm]{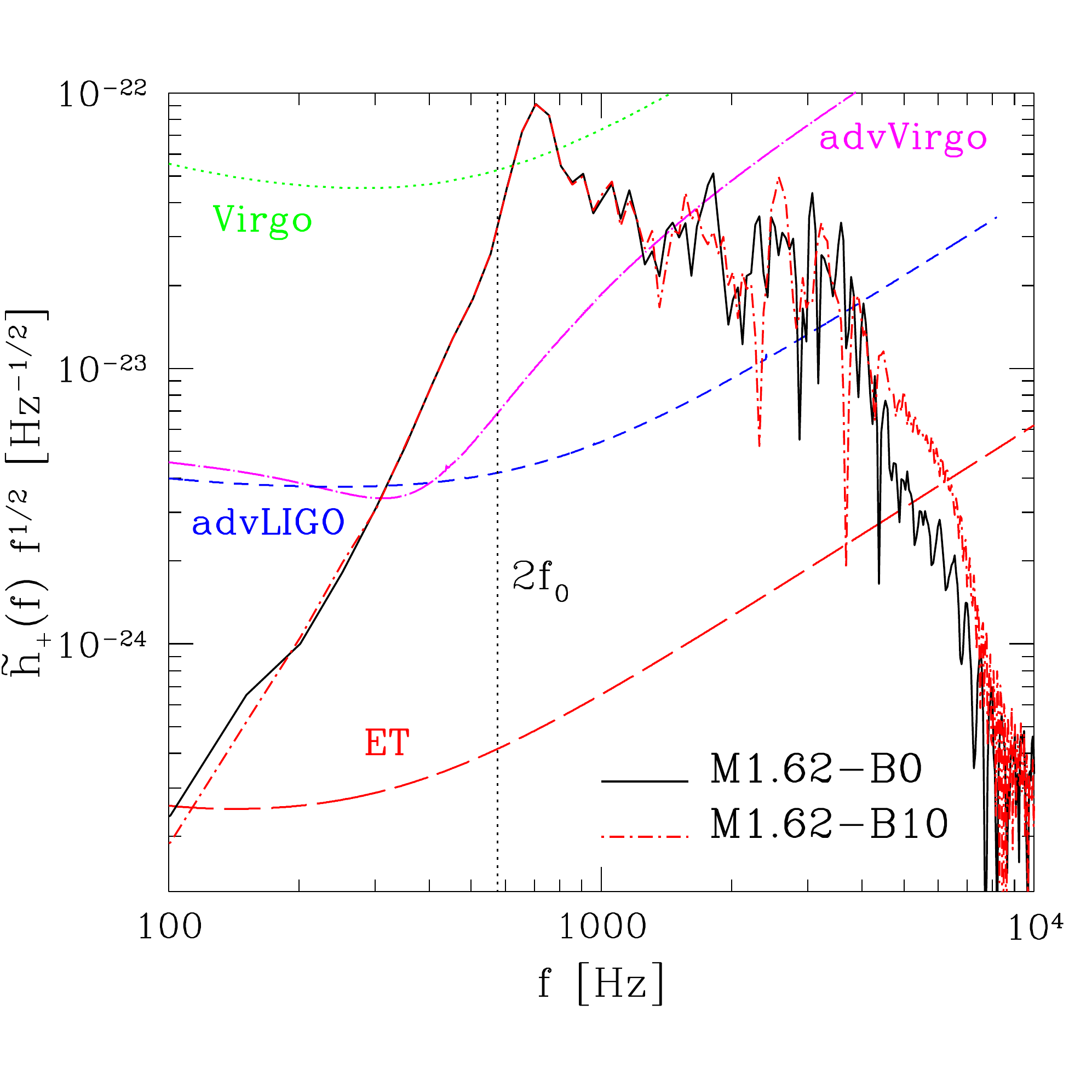} 
   \includegraphics[angle=0,width=7.0cm]{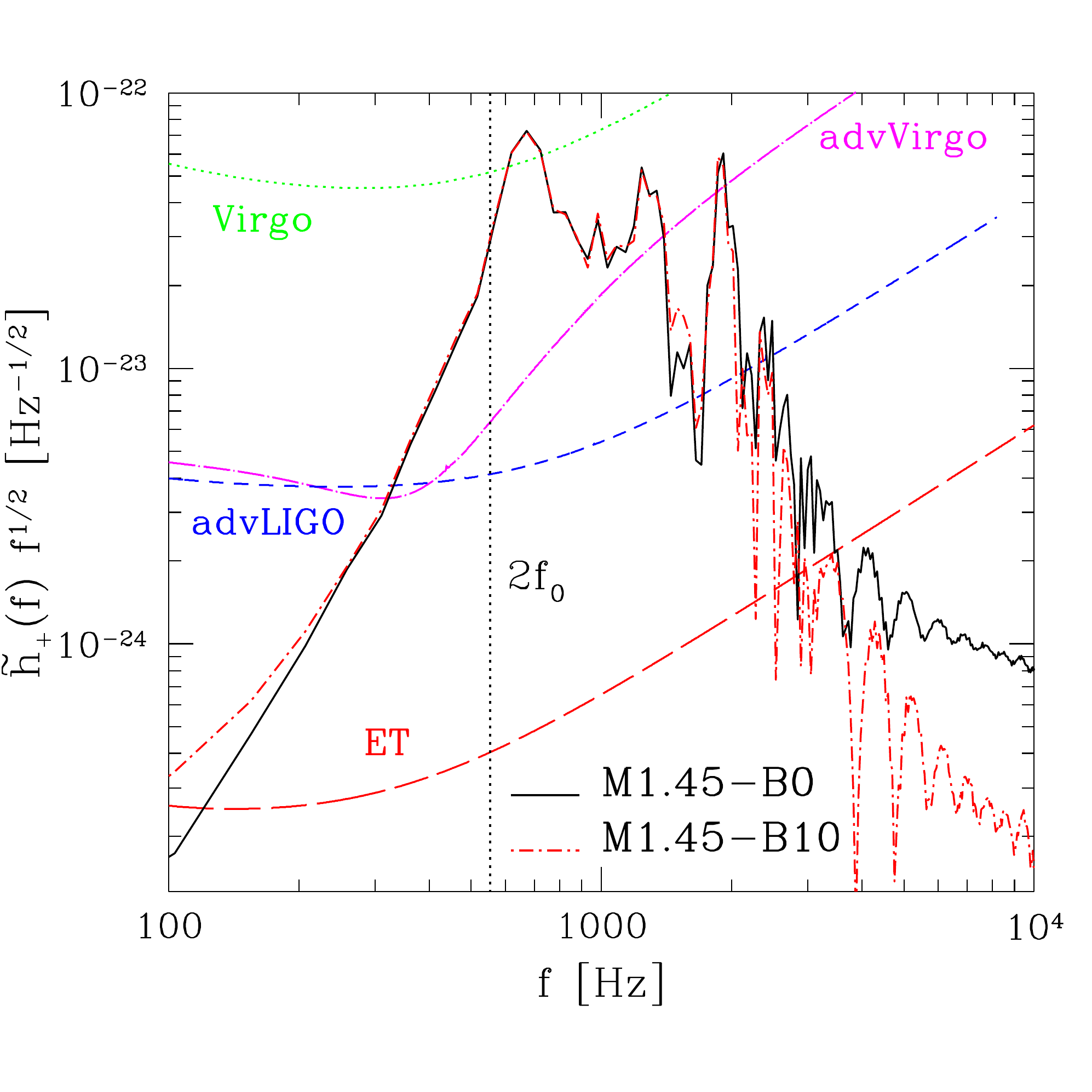}
   \includegraphics[angle=0,width=7.0cm]{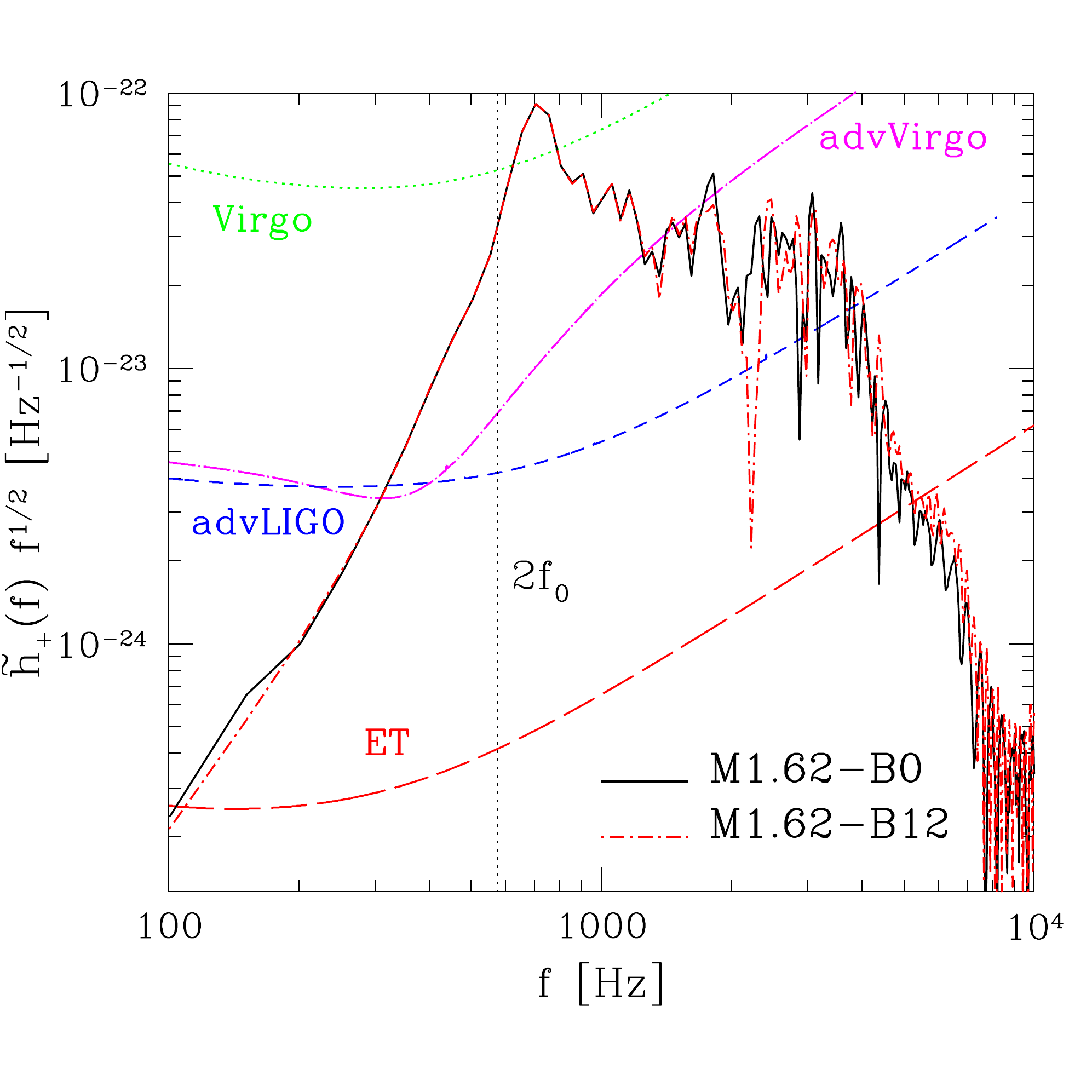} 
   \includegraphics[angle=0,width=7.0cm]{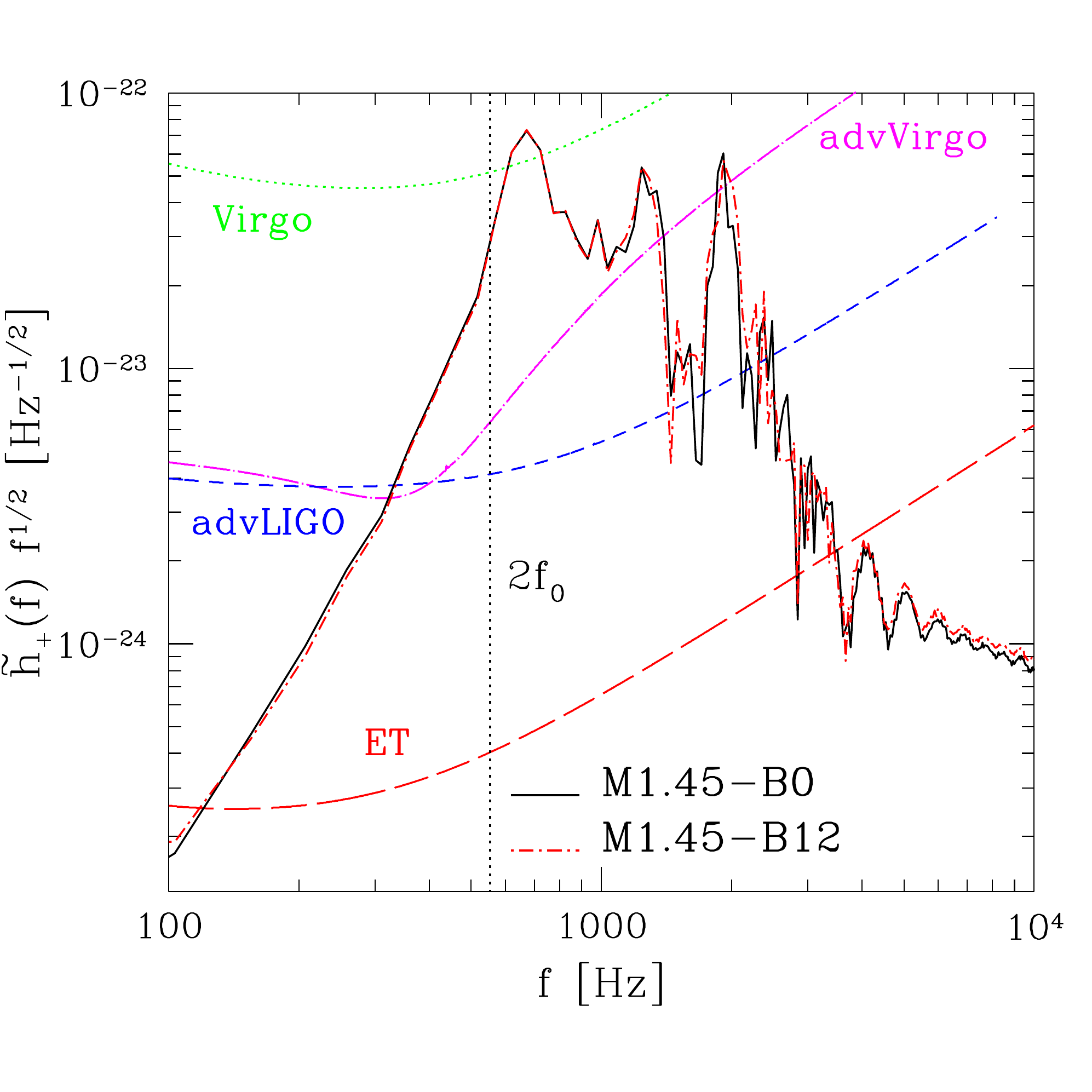}
\end{center}
    \vskip -0.5 cm
   \caption{\label{fig:PSD}Scaled power spectral densities ${\tilde
       h}_+(f) f^{1/2}$ for the high-mass case (left panel) and
     low-mass case (right panel) without magnetic field (solid black
     line) or with an initial magnetic field (dot-dashed red line) of
     $\approx 10^{8}\,{\rm G}$ (first row), $\approx 10^{10}\,{\rm G}$
     (second row) or $\approx 10^{12}\,{\rm G}$ (third row). In all
     the panels the sources are considered when placed at a distance
     of $100\,\mpc$. Shown also are the noise curves of the Virgo
     detector (dotted green line), of the advanced LIGO and advanced Virgo
     detectors (short-dashed blue and dot-dashed magenta lines,
     respectively), and of the planned Einstein Telescope (dashed red
     line). The dotted vertical lines indicate the value of twice the
     initial orbital frequency $f_0$.}
\end{figure*}

\subsection{Detectability of the magnetic field}
\label{sec:snr}

In order to assess the possibility of distinguishing between the
different waveforms and hence establish whether different
magnetizations of the HMNS can be measured, we have computed the power
spectral densities of the GWs discussed before and plotted them in
Fig.~\ref{fig:PSD} against the sensitivity curves of different
ground-based GW detectors.  In particular, we show the scaled power
spectral densities ${\tilde h}_+(f) f^{1/2}$ for the high-mass case
(left panel) and low-mass case (right panel) with an initial magnetic
field (dot-dashed red line) of $B \approx 10^{8}\,{\rm G}$ (first
row), $B \approx 10^{10}\,{\rm G}$ (second row), and $B\approx
10^{12}\,{\rm G}$ (third row). In all the panels the sources are
considered to be placed at a distance of $100\,\mpc$. We report also the
spectral densities for binaries without magnetic field (solid black
line) to aid in the comparison. Also shown are the noise curves of the
Virgo detector (dotted green line), of the advLIGO~\cite{AdvLIGO} and 
advVirgo~\cite{AdvVirgo09} detectors (short-dashed blue and dot-dashed magenta lines,
respectively), and of the planned Einstein
Telescope~\cite{Punturo:2010,ET_noise} (dashed red line). With a
dotted vertical line we indicate the value of twice the initial
orbital frequency $f_0$, so that the signal for $f < 2f_0$ should be
ignored.

In all the panels the part of the signal up to about $700\,\hz$ is
associated with the inspiral part of the waveform and in the case of
the high-mass binaries (panels in the left column) it is also the
strongest peak. The low-mass binaries (panels in the right column)
also show an additional peak with an amplitude comparable to that at
$f\approx 700\,\hz$ and it is related to (twice) the spinning period
of the bar-deformed HMNS. That peak appears for all the models at a
frequency of approximately $2\,\khz$ and its amplitude is sufficiently
high to enter into the band of advLIGO.

While we expect the position in frequency of the peak to be accurate,
its amplitude clearly depends on the subsequent evolution of the HMNS,
which we have followed here only for about $12\,\ms$.  Clearly, should
the HMNS survive on much longer timescales as shown
in~\cite{Rezzolla:2010} (see the right panel of Fig. A1 in the
Appendix of~\cite{Rezzolla:2010}), then the amplitude of this peak
could be considerably larger and could scale with the square root of the
period in which the HMNS continues to rotate before collapsing. Even
when leaving aside the role that the energy extraction via neutrinos
may play on the evolution of the post-merger object, the
hydrodynamical survival of the bar deformation in a rapidly rotating
star is still a matter of debate. The general-relativistic simulations
of isolated NSs first carried out in Ref.~\cite{Shibata:2000jt} and
then analyzed in great detail in Ref.~\cite{Baiotti06b, Manca07,
  Corvino:2010}, all indicate that the bar deformation persists only
over a timescale which is comparable with (or slightly larger than)
the dynamical one\footnote{Similar results have been found also in
  Newtonian simulations~\cite{New2000} and also for magnetized
  stars~\cite{Camarda:2009mk}}. This is due to the coupling between
the $m=2$ bar deformation with other unstable modes (most notably the
$m=1$ one), which grow to comparable amplitudes and suppress the
instability, redistributing angular momentum (see
also~\cite{Saijo2008} for a perturbative analysis in terms of a
Faraday resonance). On the other hand, simulations of stellar-core
collapse (see~\cite{Ott09} for a recent review and a complete set of
references) and the very long simulations carried out
in~\cite{Rezzolla:2010} suggest that bar-deformed stellar cores or
HMNSs can be produced and survive on timescales much longer than the
dynamical one. This different behavior in the persistence of the bar
deformation may well be due to the very different distribution of
angular momentum and density stratification between the two
configurations. Work is ongoing to confirm whether this is actually
the case.

Also quite evident from all the panels is that the spectra are very
similar but not identical and that these differences become more
appreciable for larger initial magnetic fields. Indeed the largest
differences appear for $B \simeq 10^{10}$ and, as for the accelerated
collapse discussed in Fig.~\ref{fig:delaytimes}, magnetic fields of
this strength are those that most influence the postmerger
dynamics. Once again, it is worth emphasizing that the spectra
presented here refer to a possibly too short portion of the evolution
of the HMNS and if the HMNS does survive on much longer timescales,
then the small differences shown here would become considerably more
pronounced and well within the sensitivities of advanced detectors.

In order to asses in a more quantitative way the possibility to detect
these small differences in the GWs, we have computed the
overlap between two waveforms $h_{_{\rm B1}},~h_{_{\rm B2}}$ from
binaries with initial magnetic fields ${\rm B1},~{\rm B2}$ as
\begin{equation}
\label{overlap}
\mathcal{O}[h_{_{\rm B1}}, h_{_{\rm B2}}] \equiv \frac{\langle h_{_{\rm
      B1}} | h_{_{\rm B2}} \rangle}{\sqrt{\langle h_{_{\rm B1}} |
    h_{_{\rm B1}} \rangle \langle h_{_{\rm B2}} | h_{_{\rm B2}}
    \rangle}}\,,
\end{equation}
where $\langle h_{_{\rm B1}} | h_{_{\rm B2}} \rangle$ is the scalar
product, defined as
\begin{equation}
\label{scalarproduct}
\langle h_{_{\rm B1}} | h_{_{\rm B2}} \rangle \equiv 4 \Re
	\int_0^\infty df \frac{\tilde{h}_{_{\rm B1}}(f) \tilde{h}_{_{\rm
	B2}}^*(f)}{S_h(f)}\,,
\end{equation}
and $\tilde{h}(f)$ is the Fourier transform of the GW $h(t)$ and
$S_h(f)$ is the noise power spectral density of the detector (we have
considered advLIGO here). Taking two waveforms, the closer
their overlap is to $1$, the harder will be for a detector to distinguish them.

\begin{table*}[t]
  \caption{\label{table:GWs}GW-related quantities. Column $2$ shows
    the total overlap computed for {advLIGO} between the magnetized
    models and the corresponding unmagnetized binary, while columns
    $3$ and $4$ represent the overlap computed over the inspiral and
    over the post-merger phase. Finally, columns $5-9$ show
    the SNR computed for different detectors for all the eight models
    considered here. The SNR has been obtained assuming a source at
    $100\,\mpc$.}
\begin{ruledtabular}
\begin{tabular}{lcccccccc}
Binary &
\multicolumn{1}{c}{${\cal O}$} &
\multicolumn{1}{c}{${\cal O}_{\mathrm{insp}}$} &
\multicolumn{1}{c}{${\cal O}_{\mathrm{postm}}$} &
\multicolumn{1}{c}{SNR (Virgo)}&
\multicolumn{1}{c}{SNR (LIGO)}&
\multicolumn{1}{c}{SNR (advVirgo)}&
\multicolumn{1}{c}{SNR (advLIGO)}&
\multicolumn{1}{c}{SNR (ET)}
\\
\hline
\texttt{M1.45-B0}   & $1.000$ & $1.000$ & $1.000$ & $0.33$ & $0.23$ & $1.94$ & $2.11$ & $38.90$ \\
\texttt{M1.45-B8}   & $0.997$ & $0.999$ & $0.926$ & $0.33$ & $0.23$ & $1.94$ & $2.10$ & $38.72$ \\
\texttt{M1.45-B10}  & $0.996$ & $0.999$ & $0.934$ & $0.33$ & $0.23$ & $1.94$ & $2.11$ & $38.82$ \\
\texttt{M1.45-B12}  & $0.996$ & $0.999$ & $0.899$ & $0.33$ & $0.23$ & $1.94$ & $2.11$ & $39.01$ \\
\texttt{M1.62-B0}   & $1.000$ & $1.000$ & $1.000$ & $0.36$ & $0.25$ & $2.00$ & $2.24$ & $42.57$ \\
\texttt{M1.62-B8}   & $0.998$ & $1.000$ & $0.938$ & $0.36$ & $0.25$ & $2.00$ & $2.24$ & $42.59$ \\
\texttt{M1.62-B10}  & $0.993$ & $1.000$ & $0.724$ & $0.36$ & $0.25$ & $2.00$ & $2.23$ & $42.48$ \\
\texttt{M1.62-B12}  & $0.997$ & $1.000$ & $0.893$ & $0.36$ & $0.25$ & $2.00$ & $2.24$ & $42.49$ \\
\end{tabular}
\end{ruledtabular}
\vskip -0.25cm
\end{table*}

The overlaps computed for all the magnetized binaries considered here
when compared with the corresponding non-magnetized models are
collected in Table~\ref{table:GWs}. Note that we present both the
total overlap ${\cal O}$, \ie the overlap computed over the full
time-series, and the overlaps computed over the inspiral only or the
post-merger only, \ie ${\cal O}_{\rm insp}$ and ${\cal O}_{\rm
  postm}$, respectively. Given the values in Table~\ref{table:GWs} and
since present and advanced detectors could potentially distinguish two
signals if \mbox{${\cal O}<0.995$}, it is clear that a detector such
as advLIGO or advVirgo would not be able to distinguish between a
magnetized binary and an unmagnetized one (\cf second column in the
Table). Similar considerations apply also when the overlap is computed
only over the inspiral phase (\cf third column in the Table). However,
if the overlap is computed only over the post-merger phase (\cf fourth
column in the Table) then it is evident that the differences among the
various binaries are much larger and the corresponding overlaps
considerably smaller. Hence, we conclude that a long-lived HMNS and a
detector with sufficient sensitivity at high frequencies (such as the
Einstein Telescope) could be able to measure the level of
magnetization in the progenitor NSs.

To complete the information about the GW emission from magnetized
BNSs, we have also computed the signal-to-noise-ratio (SNR) defined as
\begin{equation}
\label{eq:SNR}
\left(\frac{S}{N}\right)^2=4\int_0^\infty
        \frac{|\tilde{h}_+(f)|^2}{S_h(f)} df \; ,
\end{equation}
for different detectors and we have listed their values in
Table~\ref{table:GWs} for a source at $100\,\mpc$. Overall, it is easy
to realize that while the current Virgo and LIGO detectors
(respectively the fifth and sixth columns) would not be able to detect
these signals, SNRs larger than $1$ are obtained when considering
advLIGO and advVirgo, and even larger than $40$ in the case of the Einstein
Telescope (last column in the Table). It is worth stressing that these
SNRs should be seen as lower limits. First, the binaries are
expected to enter the sensitivity band at lower frequencies than the
ones considered here, hence adding considerable power to the
SNR. Second, as discussed extensively above, the possibility of a
long-lived HMNS could significantly add to the power at high
frequencies, hence increasing the SNR.

In summary, the results presented here indicate that BNSs do represent
strong sources of GWs and that these can be detected at distances up
to $100\,\mpc$ by the planned advanced interferometers. Determining
the level of magnetization of the progenitor stars will be very
difficult if the detected signal is confined essentially to the
inspiral, while it could be possible if the HMNS survives for
sufficiently long times as a deformed and spinning bar. In this latter
case, detectors which have high sensitivities at high frequencies,
such as advLIGO and more importantly the Einstein Telescope, will be
in a good position to measure the strength of the magnetic fields and
hence extract important physical and astrophysical information on the
progenitor NSs.

\section{Conclusions}
\label{sec:conclusions}

There is little doubt that BNSs represent prime sources for present
and advanced GW detectors. Equally clear is that NSs are observed to
have large magnetic fields, with values which can be as high as
$10^{16}\,\G$ for isolated and young magnetars. It is therefore of
great importance to assess what role the magnetic fields play
during the inspiral and merger on BNSs. Extending the research presented
in~\cite{Giacomazzo:2009mp}, we have presented the first
numerical simulations of magnetized BNSs with astrophysically
realistic magnetic fields. More specifically, we have carried out a
systematic investigation of the dynamics of both matter and magnetic
fields of equal-mass BNSs. While previous works~\cite{Anderson2008,
  Etienne08, Giacomazzo:2009mp} considered only astrophysically
unrealistic magnetic fields ($B\approx 10^{16}-10^{17}\,\G$) or
focused mainly on the inspiral part~\cite{Giacomazzo:2009mp}, here we have
considered magnetic-field values ranging from $10^{8}$ to
$10^{12}\,\G$, and evolved BNSs through all the stages of the inspiral,
merger, HMNS evolution, and collapse to BH.

Overall, we have shown that realistic magnetic fields do not affect
sensibly the dynamics of the inspiral, but they can influence that of
the post-merger, where they can accelerate the collapse of the HMNS. The
different time intervals from the merger to the collapse of the HMNS
also imply that the tori produced around the BH have slightly
different masses, reflecting the different distributions of matter and
angular momentum at the time of collapse. As a result of the tight
correlation between the degree of magnetization of the NS matter and
the delay time of the collapse, the measurement of the latter via a GW
detection will allow us to infer the former. To the best of our
knowledge, this is the first time that effects of this type have been
discussed in the evolution of inspiralling and magnetized NSs.

Magnetic fields can be amplified at the merger of the binary, when a
Kelvin-Helmholtz instability develops between the outer layers of the
two stars. Although the resolution used here is the highest
employed so far in simulating magnetized BNSs and it is sufficient to reveal
the development of the instability and the exponential growth of the
toroidal magnetic field, the amplification we have measured is only of
about one order of magnitude and is much smaller than that reported in
Ref.~\cite{Price06}, where the newly produced fields reach values in
equipartition with the kinetic energy. Although it is possible that
the different results are due to the different numerical methods
employed in Ref.~\cite{Price06}, we believe the reason behind our
modest amplifications to be that the shear layer between the two stars
survives only for about $1\,\ms$, before being destroyed by the
collision between the two stellar cores. Such a short timescale and
the relatively small velocities at the shear layer are probably
insufficient to yield the type of amplification that has been obtained
in more ``controlled'' simulations of the Kelvin-Helmholtz
instability~\cite{Zhang09,Obergaulinger10}.

The toroidal magnetic field continues to be amplified also after the
Kelvin-Helmholtz instability has been suppressed and it can reach
values that are comparable with the initial poloidal one either during
the evolution of the HMNS (in the case of low-mass binaries) or during
the evolution of the torus produced after the HMNS collapse (in the
case of high-mass binaries). This result is particularly important
since it suggests that the magnetic-field topology in the tori formed
from BNS mergers is not purely poloidal, contrarily to what has been assumed so far
by some other research groups that perform simulations of magnetized accretion disks.

When considered in terms of their GW emission, the magnetized binaries
studied here show that it is unlikely that the degree of magnetization
will be measurable by present and advanced detectors if the inspiral
is the only part of the signal available. However, if the HMNS
survives for sufficiently long times as a deformed and spinning bar,
then the modifications introduced by the presence of magnetic fields
could lead to waveforms which differ appreciably from those of
non-magnetized binaries. In this case, detectors which have high
sensitivities at frequencies larger than about $2\,\khz$, such as
advLIGO and, more importantly, the Einstein Telescope, will be able to
measure these effects for binaries up to distances of about
$100\,\mpc$.

\begin{acknowledgments}
  We thank the developers of \texttt{Lorene} for providing us with
  initial data and those of \texttt{Cactus} and \texttt{Carpet} for
  the numerical infrastructures used by \texttt{Whisky}. Useful input
  from J.~Read, C.~Reisswig, E.~Schnetter, A.~Tonita, A.~Vicer\`e, and
  S.~Yoshida is also acknowledged. We also thank M.~Koppitz for
  assisting us in the production of Fig.~\ref{fig0.5}. The
  computations were performed on the Damiana Cluster at the AEI, on
  QueenBee through LONI (\texttt{www.loni.org}), and at the Texas
  Advanced Computing Center through TERAGRID Allocation No.
  TG-MCA02N014. This work was supported in part by the DFG Grant
  SFB/Transregio~7, by ``CompStar'', a Research Networking Programme
  of the European Science Foundation, by the JSPS Grant-in-Aid for
  Scientific Research (19-07803), by the MEXT Grant-in-Aid for Young
  Scientists (22740163) and by NASA Grant No. NNX09AI75G.

\end{acknowledgments}

\bibliographystyle{apsrev-nourl}
\bibliography{aeireferences}

\end{document}